\documentclass[twocolumn]{aa}  

\usepackage{natbib}
\usepackage{orcidlink}
\usepackage{graphicx}
\usepackage{comment}
\usepackage{txfonts}
\usepackage{lipsum}
\usepackage{subcaption}         
\usepackage{lscape}             
\usepackage{placeins}           
\newcommand{\ie}{i.e., ~}
\newcommand{\eg}{e.g., ~}                              
\usepackage{hyperref}
\hypersetup{
  colorlinks=true,        
  linkcolor=blue,         
  citecolor=cyan,         
  urlcolor=cyan            
}
\begin{document}

\title{\texttt{FPIC}: a new Particle-In-Cell code for stationary and
  axisymmetric black-hole spacetimes}


%

\author{C. Meringolo \orcidlink{0000-0001-8694-3058} \inst{1}
  \and L. Rezzolla \orcidlink{0000-0002-1330-7103} \inst{1, 2, 3} }
   
\institute{Institut f\"ur Theoretische Physik, Goethe Universit\"at, 
     Max-von-Laue-Str. 1, D-60438 Frankfurt am Main, Germany
\and School of Mathematics, Trinity College, Dublin 2, Ireland
\and Frankfurt Institute for Advanced Studies, Ruth-Moufang-Str. 1, 60438
Frankfurt am Main, Germany}

   
 
\abstract
       {}
{Black holes in astrophysical environments are expected to be immersed in
  magnetized collisionless plasmas, whose electrodynamics leads to some
  of the powerful energetic emissions observed. To model such
  environments, a kinetic description is best-suited as it allows to
  capture the microphysics underlying the complex phenomena near such
  compact objects. In recent years, general-relativistic Particle-In-Cell
  (GRPIC) simulations have received increasing interest from the
  astrophysics community, becoming an excellent tool to study the
  immediate vicinity of the event horizons of stellar-mass and
  supermassive black holes. In this paper we present a newly developed
  GRPIC code framework called \texttt{FPIC}, providing a detailed
  description of the Maxwell-equations solver, of the particle
  ``pushers'', and of the other algorithms that are needed in this
  approach.}
{Special attention is paid here to the fundamental issue of
  reproducibility and hence to a detailed discussion of the numerical
  methods employed and of the potential pitfalls that can be encountered
  in their implementation. We therefore describe in detail the code,
  which is written in Fortran and exploits parallel architectures using
  MPI directives both for the fields and particles. \texttt{FPIC} adopts
  spherical Kerr-Schild coordinates, which encode the overall spherical
  topology of the problem while remaining regular at the event
  horizon. The Maxwell equations are evolved using a finite-difference
  time-domain solver with a leapfrog scheme, while multiple particle
  ``pushers'' are implemented for the evolution of the particles. In
  addition to well-known algorithms, we introduce a novel hybrid method
  that dynamically switches between the most appropriate scheme based on
  the violation of the Hamiltonian energy.}
{We first present results for neutral particles orbiting around black
  holes, both in the Schwarzschild and Kerr metrics, monitoring the
  evolution of the Hamiltonian error across different integration
  schemes. We apply our hybrid approach, showing that it is capable of
  achieving improved energy conservation at reduced computational
  cost. Results for charged particles in the presence of a magnetic field
  are also presented, where the interpolation plays a crucial role in
  maintaining Hamiltonian conservation. We then apply \texttt{FPIC} to
  investigate the Wald solution, first in electrovacuum and subsequently
  in plasma-filled configurations. In the latter case, particles with
  negative energy at infinity are present inside the ergosphere,
  indicating that the Penrose process is active. Finally, we present the
  split-monopole solution in a plasma-filled environment and successfully
  reproduce the Blandford-Znajek luminosity, finding very good agreement
  with analytical predictions.}
   {} 

\keywords{Black hole physics - High energy astrophysics -- Plasma astrophysics }

\maketitle
\nolinenumbers

\section{Introduction}
\label{sec:intro}

There is little doubt that black holes provide extreme physical
conditions and whether at the centre of galaxies -- as in the case of
active galactic nuclei (AGNs) -- or produced by the merger of neutron
stars -- as in the case of short gamma-ray bursts -- they have
long been considered responsible for the launching of powerful
relativistic jets of plasma.

In the case of AGNs, the physics behind these extreme phenomena has been
recently explored with the first images of the supermassive black holes
M87* \citep{EHT_M87_PaperI}, and Sgr A*
\citep{EHT_SgrA_PaperI}. Considerable effort has been dedicated over the
last years to the modelling, via general-relativistic
magnetohydrodynamics (GRMHD) simulations, of the accretion of plasma onto
black holes~\citep{DelZanna2020, Nathanail2022, Dihingia2022, Osorio2022,
  Das2022, Fujibayashi2020, Dhruv2025, Megale2025, Jacques2025,
  Uniyal2025}, so as to interpret the astronomical observations of the
Event Horizon Telescope Collaboration and extract physically relevant
information. GRMHD simulations have also shown that rapidly spinning
black holes threaded by a strong magnetic field can have its rotational
energy extracted electromagnetically via the Blandford-Znajek process,
launching powerful jets~\citep{Blandford1977, Tchekhovskoy2011,
  Koide2008, McKinney2012, Asenjo2017, Comisso2021, Camilloni2025}.

Despite the overall success to reproduce the large-scale features of the
accreting flow, one of the limitations of GRMHD simulations is that -- in
their most common implementation -- they treat the plasma as a single
fluid, only modelling the dynamically important part of the fluid, the
protons -- or, in general, the ions; \citep[see][for first attempts to
  two-species GRMHD simulations]{Dihingia2022, Most2022c,
  Gorard2025}. Such a description leaves completely undetermined the
microphysical properties of the electrons, such as their energy
distribution, the number densities, and temperatures. This represents a
significant limitation, since in hot, ionised jets around black holes,
the Coulomb coupling between electrons and protons is inefficient, so
that protons and electrons are likely to have distinct temperatures, as
it happens in the solar wind~\citep{Tu1997, vanderHolst2010,
  Dihingia2022, Osorio2025}. To cope with this problem, a large number of
\textit{kinetic} Particle-In-Cell approaches have been used in
special-relativistic flat spacetimes, modelling a small domain of plasmas
around compact objects, where the curvature can be
neglected~\citep{Cerutti2015, Sironi2014, Werner2018, Zhdankin2019,
  Wong2020, Sironi2021, Meringolo2023, Imbrogno2024, Imbrogno2025}.

In response to the need of having a microscopical, kinetic description of
the plasma also in regions of large curvature, such as those in the
vicinity of black holes and neutron stars, a number of
general-relativistic Particle-In-Cell (GRPIC) codes have been developed
only in the recent few years, thanks to a combination of advances in
numerical codes and computational power. The first global two-dimensional
(2D) GRPIC simulations have been presented by~\citet{Parfrey2019}, making
use of the \texttt{GRZeltron} code, which studied how a black hole
immersed in an asymptotically uniform Wald magnetic field produces jets
and extracts rotational energy according to the Blandford-Znajek
mechanism. It was also shown that a variant of the Penrose process can
also be activated, allowing particles with negative energy at infinity to
extract energy from the black hole. The same code was used for a number
of subsequent papers both in 2D~\citep{Crinquand2020, Crinquand2020b,
  Bransgrove2021, ElMellah2021, Galishnikova23, Niv2023, Vos2024} and in
three-dimensional (3D) scenarios~\citep{Crinquand2022, ElMellah2023,
  Figueiredo2025}. All of these studies have highlighted the richness of
the plasma dynamics near rotating black holes and the effectiveness of
GRPIC simulations to model processes such as the disc accretion of
magnetically dominated flows and the powering of relativistic jets.

Another GRPIC code that has been recently developed and is publicly
available is \texttt{Aperture}~\citep{Chen2025}, which was applied to
study spark gaps and plasma injection in black hole magnetospheres,
finding that the location and time evolution of the gap depend on the
observer. Further works with this code are the self-consistent treatment
of inverse Compton scattering with the inclusion of pair
production~\citep{Yuan2025}, and the study of kinetic equilibria of
collisionless tori around a Kerr black hole that are stable in 2D
axisymmetric simulations in the absence of and initial seed magnetic
field~\citep{Luepker2025}. Additional works in this direction are those
presented by~\citet{Hirotani2021}, who developed a simplified GRPIC
approach evolving only three electromagnetic-field components to study an
axisymmetric magnetosphere around a rotating black hole and the recent
contribution by~\citet{Galishnikova2025arXiv}, who presented the
open-source, coordinate-agnostic GRPIC code \texttt{Entity}.

In this paper, we present a newly developed GRPIC code, \texttt{FPIC},
for stationary and axisymmetric black hole spacetimes. One of the major
goals of this paper is to ensure reproducibility of scientific
results. As a result, special attention is paid to a detailed discussion
of the numerical methods employed in \texttt{FPIC}, and to the discussion
of the potential difficulties that can be encountered when developing a
GRPIC code from scratch.  However, this is not a code-description paper
only. In addition to the discussion of well-known numerical aspects and
of standard code benchmarks, we also present a novel and more efficient
method for the evolution of the charged particles and an astrophysical
application to the study of the Blandford-Znajek process in a
split-monopole magnetic topology.

The paper is organised as follows. In Section~\ref{Sec2}, we introduce
the key ingredients of a GRPIC code and describe the main algorithms
implemented in \texttt{FPIC}. In Section~\ref{Sec3}, we present a series
of simulations in order to validate the code. We begin with
single-particle trajectories around Schwarzschild and Kerr black holes,
introducing a new hybrid integrator. We then present results for rotating
black holes immersed in Wald and split-monopole magnetic-field
configurations. Finally, in Section~\ref{SecCon}, we summarise the
results and discuss the main properties of the code. Hereafter, we employ
Greek letters for tensor components in a four-dimensional manifold and
Latin letters for the corresponding spatial components. We also adopt the
summation convention on repeated indices and geometrised
units~\citep[see, \eg][]{Rezzolla_book:2013}.

\section{The GRPIC algorithm}
\label{Sec2}

\texttt{FPIC} models the dynamics of charged particles in a stationary
and axisymmetric spacetime, described by the black hole mass $M=1$ and by
the dimensionless spin parameter $0 \leq a_* \leq 1$, being $a_* :=
J_*/M^2$ and $J_*$ the spin angular momentum of the spacetime. The code
employs the standard $3+1$ spacetime decomposition
\citep{Thorne82, Komissarov2004b}, where the line element is described by:
\begin{equation}
ds^2 = -\alpha^2 dt^2 + \gamma_{ij}(dx^i + \beta^i dt)(dx^j + \beta^j dt)\,.
\end{equation} 
As usual, the three-metric is represented by $\gamma_{ij}$, while
$\alpha$ and $\beta^i$ refer to the lapse function and the shift vector, 
respectively \citep{Rezzolla_book:2013}. To ensure regularity at the
black hole horizon, the metric is expressed in terms of the spherical
Kerr-Schild coordinates $(t, r, \theta, \varphi)$, because these encode
the spherical symmetry of the problem and remain regular at the event
horizon (see Appendix~\ref{AppKS} for details). All the fields are
invariant by rotation around the spin axis of the black hole, \ie
$\partial_\varphi \psi(r, \theta)=0$ for all quantities $\psi$ in the
code. In that sense, although particles move in a full 3D space, the
fields are defined by 2D arrays, being often called ``2.5D''
approximation. Hereafter, and unless specified differently, we use
geometrised units in which $c = 1 = G$, with $c$ and $G$ the speed of
light and the gravitational constant, respectively.

\subsection{Maxwell Equations Solver}

\begin{figure}
\centering
\includegraphics[width=0.8\columnwidth]{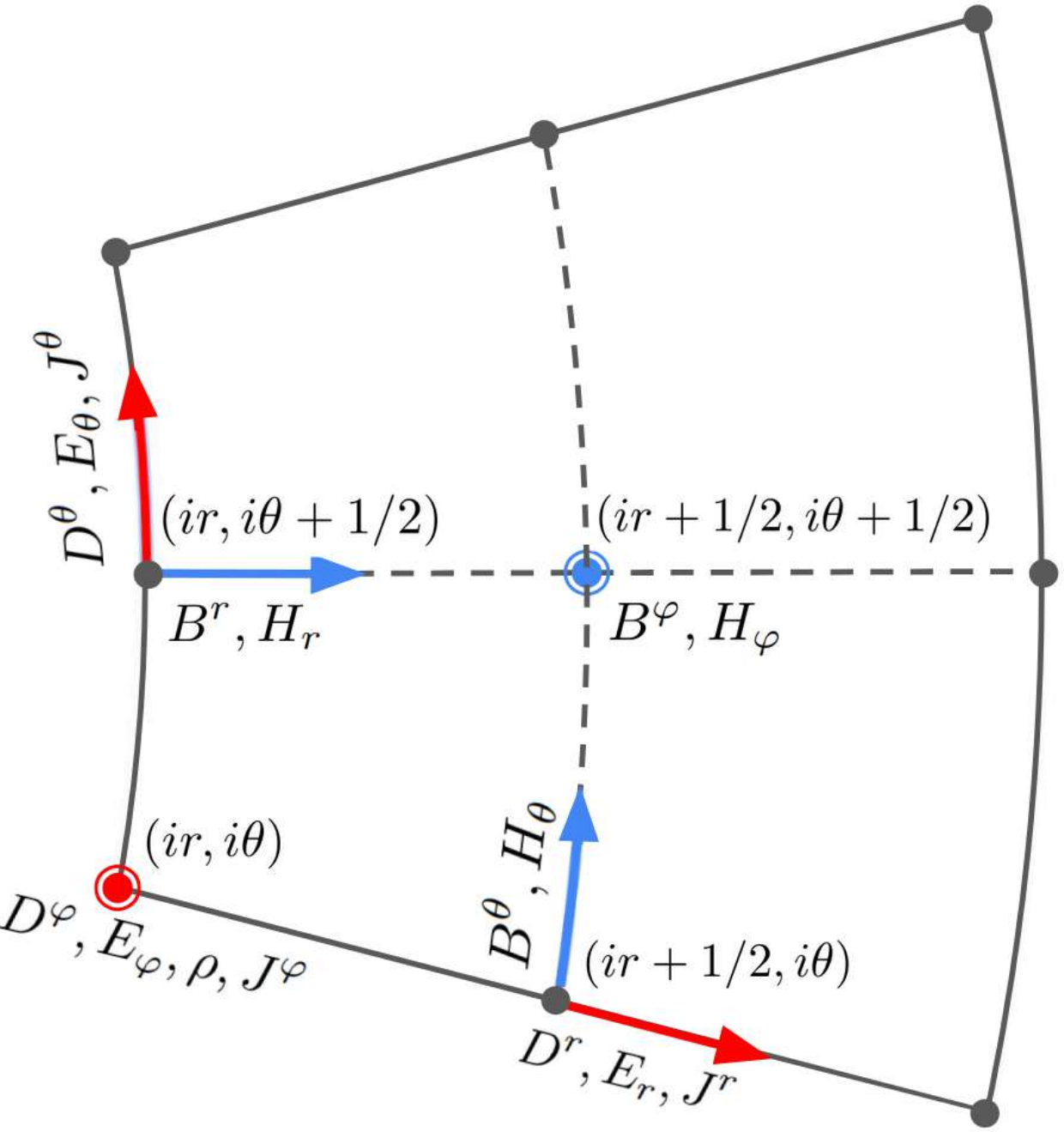}
\caption{Spherical Yee grid adopted in \texttt{FPIC}. The red arrows
  denote the components of the electric fields $\boldsymbol{D}, 
  \boldsymbol{E}$, and the current density $\boldsymbol{J}$, the blue
  ones denote the magnetic fields $\boldsymbol{B}, \boldsymbol{H}$. The
  positions on the mesh are labelled by the indices $ir$ in the radial
  direction and $i\theta$ in the radial direction. Note that at
  $(ir, i\theta)$ and ($ir+1/2, i\theta+1/2$) the vectors are pointing
  outwards.}
\label{fig:1} 
\end{figure}

For static electromagnetic fields, the covariant antisymmetric Maxwell
tensor can be expressed in terms of derivatives of the four-potential
$A_\mu$, as
\begin{equation}
F_{\mu \nu} = g_{\mu \alpha} g_{\nu \beta} F^{\alpha \beta} 
= \partial_\mu A_\nu - \partial_\nu A_\mu \,.
\end{equation}
\texttt{FPIC} evolves the contravariant electric field $D^i$ and magnetic
field $B^i$ as dynamic variables as measured by a fiducial observer
(FIDO), defined by~\citep{Komissarov2004b, Mizuno2024}
\begin{equation}
D^i = \alpha F^{0i}\,, ~~~~ B^i = \alpha\, {}^{*}\!F^{i0} \,, 
\end{equation}
where we have defined the dual of the Maxwell tensor
\begin{equation}
{}^{*}\!F_{\mu \nu} := \sqrt{-g} \, \eta_{\mu \nu \alpha \beta} F^{\alpha \beta} \,, 
\end{equation}
being $\eta_{\nu \mu \alpha \beta}$ the completely anti-symmetric
Levi-Civita symbol. The general-relativistic Maxwell equations in a
stationary metric can be written as \citep{Baumgarte2010a,
  Rezzolla_book:2013}:
\begin{align}
\dfrac{1}{\sqrt{\gamma}} \partial_j \big( \sqrt{\gamma} D^j \big)
  &= 4 \pi \rho \,, 
\label{m1}
\\
\dfrac{1}{\sqrt{\gamma}} \partial_j \big( \sqrt{\gamma} B^j \big)
  &= 0 \,, 
\label{m2}
\\
\dfrac{1}{\sqrt{\gamma}} \eta^{ijk} \partial_j H_k
  - 4 \pi J^i
  &= \partial_t D^i \,, 
\label{m3}
\\
-\dfrac{1}{\sqrt{\gamma}} \eta^{ijk} \partial_j E_k
  &= \partial_t B^i \,, 
\label{m4}
\end{align}
where $\boldsymbol{J} := \alpha \boldsymbol{j} - \rho
\boldsymbol{\beta}$, $\boldsymbol{j}$ is the current density as measured
by FIDOs and $\eta_{ijk}= \eta^{ijk}$ is the 3-dimensional Levi-Civita
symbol. Note that Eqs.~(\ref{m1})--(\ref{m2}) are simply constraints and
are not evolved during the simulations. The evolution equations
(\ref{m3}) and (\ref{m4}) refer to FIDO quantities but make use of the
electric field $\boldsymbol{E}$ and the magnetic field $\boldsymbol{H}$
as measured from the grid, where:
\begin{align}
H_i &= \dfrac{1}{2}\alpha \sqrt{\gamma} \, \eta_{ijk}F^{jk} = \alpha B_i - 
\sqrt{\gamma} \, \eta_{ijk} \, \beta^j D^k , 
\\
E_i &= F_{i0} = \alpha D_i + \sqrt{\gamma} \, \eta_{ijk} \, \beta^j B^k .
\label{Ei}
\end{align}
We make use of the Yee grid~\citep{Yee66}, where all the electromagnetic
fields are staggered and located in different grid positions
\citep{Cerutti2015, Torres2024}, as sketched in Fig.~\ref{fig:1}. 
One of the major benefits of employing the Yee grid is that Eq.~(\ref{m2}) is
automatically satisfied to machine round-off precision, provided the
simulation is initialised with a divergence-less magnetic field, \ie
$\partial_t \big[\partial_j \big( \sqrt{\gamma} B^j \big)/\sqrt{\gamma}
  \big]= 0$.

The code solves the time-dependent Maxwell equations (\ref{m3}) and
(\ref{m4}) via a time-centered leapfrog scheme
\citep{iserles1986generalized, birdsall2005plasma}. Electric and magnetic
fields are staggered not only in space, but also in time. The code knows
$\boldsymbol{B}$ and $\boldsymbol{H}$ at integer time steps $n$ and
$n+1$, while electric fields $\boldsymbol{D}$ and $\boldsymbol{E}$ are
stored at semi-integers timesteps $n+1/2$ and $n+3/2$. In order to evolve
electromagnetic fields and particles, the following sub-steps are
employed~\citep{Crinquandthesis, Galishnikova2025arXiv}:
\begin{enumerate}
    \item Compute ${\boldsymbol{E}}^{n+1} = \alpha \langle
      {\boldsymbol{D}} \rangle^{n+1} + \boldsymbol{\beta} \times
      {\boldsymbol{B}}^{n+1}$\, ;
    \vspace{8pt}
    \item Evolve $\langle {\boldsymbol{B}}\rangle^{n+1/2}$ to
      ${\boldsymbol{B}}^{n+3/2}$ via Eq.~(\ref{m4}) and
      ${\boldsymbol{E}}^{n+1}$\, ;
    \vspace{8pt}
    \item Compute ${\boldsymbol{E}}^{n+3/2} = \alpha
      {\boldsymbol{D}}^{n+3/2} + {\boldsymbol{\beta}} \times
      {\boldsymbol{B}}^{n+3/2}$\, ;
    \vspace{8pt}
    \item Evolve ${\boldsymbol{B}}^{n+1}$ to ${\boldsymbol{B}}^{n+2}$ via
      Eq.~(\ref{m4}) and ${\boldsymbol{E}}^{n+3/2}$\, ;
    \vspace{8pt}
    \item Push particles at ${\boldsymbol{x}}^{n+1}$ and four-velocity
      ${\boldsymbol{u}}^{n+1}$ to $({\boldsymbol{x, u}})^{n+2}$ with
      ${\boldsymbol{B}}^{n+3/2}, {\boldsymbol{D}}^{n+3/2}$ \, ;
    \vspace{8pt}
    \item Update source terms $\rho^{n+2}, {\boldsymbol{J}}^{n+2}$\, ;
    \vspace{8pt}
    \item Compute ${\boldsymbol{H}}^{n+3/2} = \alpha
      {\boldsymbol{B}}^{n+3/2} - {\boldsymbol{\beta}} \times
      {\boldsymbol{D}}^{n+3/2}$\, ;
    \vspace{8pt}
    \item Evolve ${\boldsymbol{D}}^{n+1}$ to ${\boldsymbol{D}}^{n+2}$ via
      Eq.~(\ref{m3}) and ${\boldsymbol{H}}^{n+3/2}, \langle
      {\rho}\rangle^{n+3/2}, \langle {\boldsymbol{J}}\rangle^{n+3/2}$\, ;
    \vspace{8pt}
    \item Compute ${\boldsymbol{H}}^{n+2} = \alpha {\boldsymbol{B}}^{n+2}
      - {\boldsymbol{\beta}} \times {\boldsymbol{D}}^{n+2}$\, ;
    \vspace{8pt}
    \item Evolve ${\boldsymbol{D}}^{n+3/2}$ to ${\boldsymbol{D}}^{n+5/2}$
      via Eq.~(\ref{m3}) and ${\boldsymbol{H}}^{n+2}, 
      {\boldsymbol{J}}^{n+2}$\,, 
\end{enumerate}
where $\langle \psi \rangle^n = (\psi^{n-1/2}+\psi^{n+1/2})/2$ indicates
an average between different timesteps. Note that, to preserve the
leapfrog scheme, we need to 1) store the fields $D^i, B^i, J^i$ and
$\rho$ at two different timesteps, and 2) double the ordinary leapfrog
pushes, in order to evolve the fields. In addition to this, a number of
other numerical subtleties are hidden when coupling FIDOs and non-FIDOs
quantities living in different Yee-grid points. To see this, we can
consider, for example, the radial component of Eq. (\ref{m3}), which
reads $\partial_t D^r = (\partial_\theta H_\varphi)/\sqrt{\gamma} - 4 \pi
J^r$. To solve this equation, we need $H_\varphi = \alpha B_\varphi -
\sqrt{\gamma} \, \beta^r D^\theta$, but note from Fig.~\ref{fig:1} that
$B_\varphi$ and $D^\theta$ are located at different grid positions. Also, 
we need to compute the covariant component $B_\varphi = \gamma_{i
  \varphi} B^i$, where again, different components need to be centered in
space. As an example, to obtain the $B^r$ component at the same grid
position of the $B^\varphi$ component, i.e. at $(ir+1/2, i\theta+1/2)$, 
we perform the following metric-weighted linear interpolation:
\begin{equation}
B^r_{(ir+1/2, i\theta+1/2)} = \dfrac{\left( \sqrt{\gamma} B^r
  \right)_{(ir, i\theta+1/2)} + \left( \sqrt{\gamma} B^r \right)_{(ir+1, 
    i\theta+1/2)}}{2 \sqrt{\gamma}_{(ir+1/2, i\theta+1/2)} }\,.
\end{equation}
Since such hidden calculations are ubiquitous for the evolution of
electromagnetic fields, all the stationary metric components ($\alpha, 
\beta^i, \gamma_{ij}$ and $\sqrt{\gamma}$) are computed only initially
and stored in all the different Yee grid points.

\subsection{Particle Pusher}

\texttt{FPIC} evolves the contravariant three-coordinate $x^i$ and the
covariant momentum vector $u_i$ as dynamic variables for the particles.
In the $3+1$ formalism and stationary metrics, the Hamiltonian for a
charged particle under the electromagnetic and gravitational forces reads
\begin{equation}
H(x^i, u_j) = \alpha \Gamma - \beta^k u_k - \frac{q}{m} A_0\,, 
\label{Ham}
\end{equation}
where $\Gamma := \sqrt{1+\gamma^{ij}u_i u_j}$ is the particle Lorentz
factor measured by a FIDO, $m$ is the mass of the particle, $q=\pm e$ the
charge, and $A_0$ is the time-component of the four-vector potential
$A_\mu$. The general-relativistic equations of motion for the position
and the four-velocity of the particles can be obtained by differentiating
$H$ such that~\citep{Bacchini2019}:
\begin{align}
\frac{d x^i}{dt} =& \frac{\partial H}{\partial u_i}
=\frac{\alpha}{\Gamma} \gamma^{ij} u_j - \beta^i\,, 
\label{p1}
\\
\frac{d u_i}{dt} =& -\frac{\partial H}{\partial x^i} =
\mathcal{A}^{\text{(g)}}_i + \mathcal{A}^{\text{(em)}}_i, 
\label{p2}
\end{align}
where the two acceleration terms in the evolution of momentum can be written as:
\begin{align}
\mathcal{A}^{\text{(g)}}_i &= -\Gamma \partial_i \alpha + u_j \partial_i
\beta^j - \frac{\alpha}{2 \Gamma} u_l u_m \partial_i \gamma^{lm}, 
\label{O1}
\\
\mathcal{A}^{\text{(em)}}_i &= \frac{q \alpha}{m} \left[ \gamma_{ij} D^j
  + \frac{\sqrt{\gamma} \, \eta_{ijk} \gamma^{jl} u_l B^k}{\Gamma}
  \right].
\label{O2}
\end{align}
To increase the accuracy of the pusher, both the metric components in
Eqs.~\eqref{p1}--\eqref{O2} and their derivatives are computed
analytically at the position of the particle.

\texttt{FPIC} implements different schemes to solve the above evolution
equations for particles: 1) an explicit fourth-order Runge-Kutta (RK4)
integrator~\citep{Press1992}; 2) an implicit midpoint rule (IMR)
\citep{Feng1986}; 3) a second-order implicit Hamiltonian method
\citep{Bacchini2018}. The two implicit methods, \ie the IMR and the
Hamiltonian methods, are the most accurate ones, but also computationally
demanding, while the RK4 scheme shows acceptable results in terms of
accuracy at a fraction of the computational cost. As a result, although
the RK4 integrator is not symplectic, \ie the area of a given region of
the phase space is not guaranteed to be preserved over long timescales, 
(as required by Liouville's theorem~\citep{Rezzolla_book:2013}), it has
represented for us the optimal approach in terms of accuracy and
computational costs for most of the cases we considered.

Once the updated positions and velocities are computed, the new charge
and current density are then moved from the sources to the Yee grid via a
metric-weighted 2D linear interpolation, as discussed in detail in the
next Subsections.

\subsubsection{Runge-Kutta 4 integrator}

The fourth-order RK4 scheme is a classical explicit integrator commonly
used in scientific community for the numerical solution of ordinary
differential equations (ODEs)~\citep{Press1992}. A key advantage of this
method is that it updates the solution through a fixed sequence of
non-iterative stages. Considering a generic function
$f^n({\boldsymbol{x}}, t)$, representing the time-discretized form of an
ODE problem at time $n$, we can write:
\begin{equation}
\dfrac{f^{n+1}-f^n}{\Delta t} = \mathcal{F}(t, f)\,, 
\end{equation}
where $\mathcal{F}(t, f)$ is a given function. For each timestep, the RK4
scheme corresponds to first recursively calculating the following
quantities
\begin{eqnarray}
k_1 &=& \Delta t \, \mathcal{F}(t^n, f^n)\,, 
\\
k_2 &=& \Delta t \, \mathcal{F}(t^n+\Delta t/2, f^n + k_1/2)\,, 
\\
k_3 &=& \Delta t \, \mathcal{F}(t^n+\Delta t/2, f^n + k_2/2)\,, 
\\
k_4 &=& \Delta t \, \mathcal{F}(t^n+\Delta t, f^n + k_3)\,, 
\end{eqnarray}
and finally advances the variable $f(t)$ by means of
\begin{equation}
f^{n+1} = f^n + \dfrac{1}{6} \left( k_1 + 2k_2 + 2k_3 + k_4 \right)\,.
\end{equation}
Notice that the RK4 scheme requires four evaluations of the sources to
advance one timestep, and the resulting error in $f(t)$ in the computed
solution is of order $\mathcal{O}(\Delta t^5)$. The main drawback of
explicit methods, such as the RK4 scheme, is that they are generally
unable to preserve the first integrals of motion, when these
exist. Moreover, explicit integrators do not conserve the phase space
volume, in other words, they are non-symplectic schemes~\citep{HaiLW02}.
As a consequence, in addition to the truncation errors of a given order
in the solution variable $f(t)$, other errors may inevitably accumulate
and grow over long-time integrations.

However, the scaling of the errors with decreasing timestep $\Delta t$ is
satisfactory enough to be generally acceptable for the RK4 method. As we
will show in the following Sections, this scheme provides the best
compromise between computational cost and accuracy, while preserving
energy conservation within acceptable values over the typical time scales
of GRPIC simulations. Further improvements for the RK4 method can be
obtained by adopting adaptive timestep control, but they can increase the
overall computational cost of the simulation, and we usually adopt a
fixed timestep unless specified differently.

\subsubsection{Implicit Midpoint Rule}

The implicit midpoint rule (IMR) is the simplest symplectic integrator, 
and it is second-order accurate~\citep{Feng1986, Bacchini2018}. It solves
ordinary differential equations $\partial_t y = f(y)$ according to
\begin{equation}
\dfrac{f^{n+1}-f^n}{\Delta t} = \mathcal{F}(t^{n+1/2},f^{n+1/2})\,, 
\label{imr1}
\end{equation}

where in the right hand side we estimate $f^{n+1/2} = (f^{n}+y^{n+1})/2$.
Note that this is an implicit method, since $f^{n+1}$ is our unknown, but
it appears on both sides of Eq.~(\ref{imr1}). We solve iteratively the
above equation, including both the terms $\mathcal{A}^{\text{(em)}}_i$
and $\mathcal{A}^{\text{(g)}}_i$ for charged particles. We update
$f^{n+1}$ for the $k$-th iteration according to
\begin{equation}
f^{n+1, [k]} = f^n + \dfrac{\Delta t }{2} \mathcal{F}\left(f^n + f^{n+1, [k-1]} \right)\,.
\label{imr2}
\end{equation}
We employ a classical Newton iterative scheme~\citep{Press1992}, which
usually converges to the machine round-off after $k \sim 10-15$ iterations
for all the six components $(x^i, u_j)$. The IMR has a higher cost than
the RK4 integrator, but it has the benefits of unconditional stability
and of being symplectic. The Hamiltonian energy is not conserved exactly, but
the error is bounded in time and of order $\mathcal{O}(\Delta
t^3)$. Moreover, the implicit nature of this scheme allows for larger
$\Delta t$ without compromising stability~\citep{birdsall2005plasma}.

\subsubsection{Hamiltonian integrator}

In some particular cases, one may need a scheme that does exactly 
conserve energy, such as the second-order, Hamiltonian-preserving
scheme~\citep{Bacchini2018, Bacchini2019}. It is straightforward to show that, 
in the continuous case, the variation of the Hamiltonian vanishes 
by application of the chain rule
\begin{eqnarray}
\dfrac{\partial H(x^i, u_j)}{dt} = \dfrac{\partial H}{\partial x^i}
\dfrac{dx^i}{dt} + \dfrac{\partial H}{\partial u_j} \dfrac{du_j}{dt} =
-\dfrac{du_i dx^i}{dt} + \dfrac{dx^i du_i}{d t} =0\,.
\label{ham0}
\end{eqnarray}
The discretized formulation of Eq.~\eqref{ham0} accordingly reads
\begin{eqnarray}
\dfrac{\Delta H}{\Delta u_i} = \dfrac{\Delta x^i}{\Delta t}\,, ~~~
\dfrac{\Delta H}{\Delta x^i} = - \dfrac{\Delta u_i}{\Delta t}\,, 
\label{hh}
\end{eqnarray}
and will in principle satisfy $\Delta H/\Delta t=0$ under certain
conditions. Since the Hamiltonian is a function of six variables, we
must treat carefully the differentiation with respect to each
variable. Here we follow the approach of~\citet{Bacchini2018}, defining
$\mathcal{H}$ the Hamiltonian containing some average relative to the
variables that are not differentiated, rewriting Eqs.~(\ref{hh}) as
\begin{eqnarray}
\dfrac{x^{i, n+1} - x^{i, n}}{\Delta t} &=&
\dfrac{\mathcal{H}({\boldsymbol{x}}, u_i^{n+1}, u_l, u_m) -
  \mathcal{H}({\boldsymbol{x}}, u_i^n, u_l, u_m)}{u_i^{n+1} - u_i^n}\,, 
\\
\dfrac{u^{n+1}_i - u^n_i}{\Delta t} &=& \dfrac{\mathcal{H}(x^{i, n+1}, 
  x^l, x^m, {\boldsymbol{u}}) - \mathcal{H}(x^{i, n}, x^l, x^m, 
  {\boldsymbol{u}})}{x^{i, n+1} - x^{i, n}}\,, 
\end{eqnarray}
and reporting the details in Appendix~\ref{ham_eq}. Note that this scheme
is completely generic and can be applied to different physical situations
provided that the corresponding Hamiltonian $H(x^i, u_j)$ is available, 
and retaining first integrals conservation properties. However, despite
this pusher is the most accurate among all the ones implemented, it
remains unfeasible when evolving a large number of particles as expected
by a PIC code, because of its high computational cost. In \texttt{FPIC}
we implemented the Hamiltonian formalism only for the geodesic motion, as
presented in~\citet{Bacchini2018}, remaining overall a very good particle
pusher for a few test-particle cases and testbeds, when only the
$\mathcal{A}^{\text{(g)}}_i$ term is present.

\subsection{Current and charge deposition}

%
\begin{figure}[t]
\centering
\includegraphics[width=0.8\columnwidth]{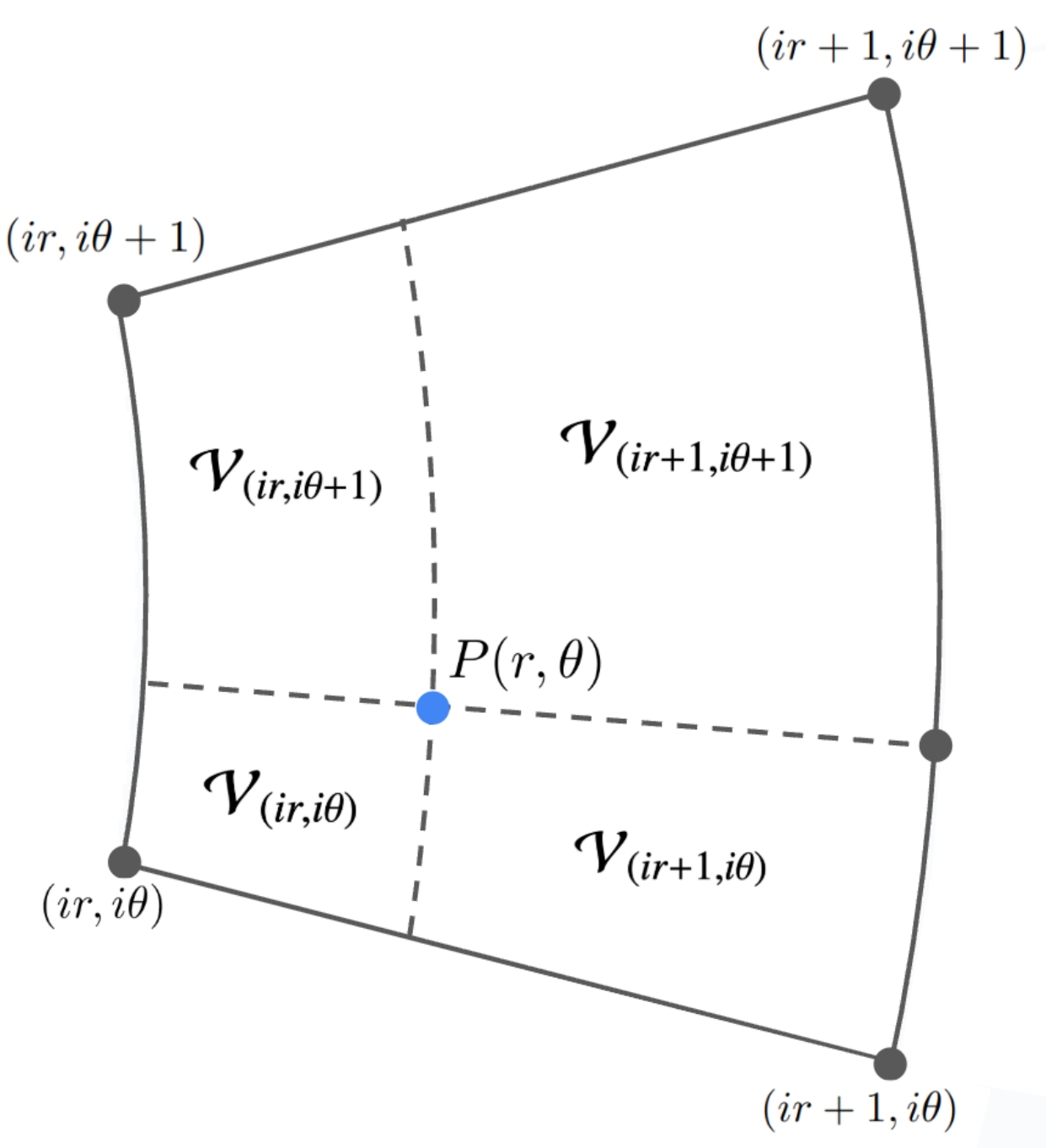}
\caption{Volume weighting procedure utilised in \texttt{FPIC}. Reported
  in the diagram is the geometry of a single cell in a 2D axisymmetric
  spherical mesh, with a particle located in the cell at position
  $P(r, \theta)$ (blue point). The volumes $\mathcal{V}$ involved in the
  interpolation scheme are also reported, accordingly.}
\label{charge} 
\end{figure}

The deposition of currents and charges is an important step for PIC
codes. To update the Amp\`ere-Maxwell law in Eq.~(\ref{m3}), the source
terms $\rho$ and $\boldsymbol{J}$ are needed, capturing the coupling
between the electromagnetic fields and the charged
particles. \texttt{FPIC} uses a 2D linear interpolation to deposit the
charge from the particles into each grid node. The proportion of a charge
which is deposited on a given node is determined by the volume defined by
the position of the particle between the four grid points of the cell.

It is worth mentioning that in PIC simulations each particle is assigned
a numerical weight $w$, which compensates for the low number of numerical
particles used in kinetic simulations compared to the actual number of
particles in real astrophysical plasmas~\citep{birdsall2005plasma}. As a
result, a single numerical particle represents a large number of physical
particles with the same charge-to-mass ratio and following identical
phase-space trajectories (for this reason, they are also referred to as
\textit{macroparticles}).

Consider a single particle located at the point $P(r, \theta)$ (see the
blue point in Fig.~\ref{charge})), which has charge $q$ and weight $w$, 
and is located between nodes $r_{ir} \leq r \leq r_{ir+1}$ and
$\theta_{i\theta} \leq \theta \leq \theta_{i\theta+1}$. The charge
deposited on each node is computed as
\begin{eqnarray}
q_{(ir, i\theta)} &=&\dfrac{\mathcal{V}_{(ir+1, i\theta+1)}}{\mathcal{V}}qw\,, 
\\
q_{(ir+1, i\theta)} &=&\dfrac{\mathcal{V}_{(ir, i\theta+1)}}{\mathcal{V}}qw\,, 
\\
q_{(ir, i\theta+1)} &=&\dfrac{\mathcal{V}_{(ir+1, i\theta)}}{\mathcal{V}}qw\,, 
\\
q_{(ir+1, i\theta+1)} &=&\dfrac{\mathcal{V}_{(ir, i\theta)}}{\mathcal{V}}qw\,, 
\end{eqnarray}
where $\mathcal{V}$ is the total volume between the four grid nodes, given by
\begin{equation}
\mathcal{V} = \int_{r_{ir}}^{r_{ir+1}} \int_{\theta_{i\theta}}^{\theta_{i\theta+1}} 
\sqrt{\gamma} \, dr d\theta\,.
\label{v1}
\end{equation}
The other \textit{subvolumes} in the cell are computed according to
\begin{eqnarray}
\mathcal{V}_{(ir, i\theta)} &=& \int_{r_{ir}}^{r} \int_{\theta_{i\theta}}^{\theta} 
\sqrt{\gamma} \, dr d\theta\,, 
\\
\mathcal{V}_{(ir+1, i\theta)} &=& \int_{r}^{r_{ir+1}} \int_{\theta_{i\theta}}^{\theta} 
\sqrt{\gamma} \, dr d\theta\,, 
\\
\mathcal{V}_{(ir, i\theta+1)} &=& \int_{r_{ir}}^{r} \int_{\theta}^{i\theta+1} 
\sqrt{\gamma} \, dr d\theta\,, 
\\
\mathcal{V}_{(ir+1, i\theta+1)} &=& \int_{r}^{r_{ir+1}} \int_{\theta}^{i\theta+1} 
\sqrt{\gamma} \, dr d\theta\,.
\label{v5}
\end{eqnarray}
Note that, even if the geometry is 2D, the determinant of the metric in
Eqs.~(\ref{v1})--(\ref{v5}) is fully 3D, and is reported in
Eq.~(\ref{det}). We evaluate the different volumes via numerical
integration by using the 2D trapezoidal rule, and discretising each cell
into $4\times 4$ points in $r$ and $\theta$.

The total charge density $\rho$ and current density $\boldsymbol{J}$
deposited at the grid point $(ir, i\theta)$, for instance, reads:
\begin{eqnarray}
\rho_{(ir, i\theta)} &=& \sum_k \dfrac{w^k \, q^k_{(ir, i\theta)}
  \mathcal{V}^k_{(ir+1, i\theta+1)}}{\mathcal{V}^k} \,, \\
{\boldsymbol{J}}_{(ir, i\theta)} &=& \sum_k \dfrac{w^k q^k_{(ir, i\theta)}
  \, {\boldsymbol{v}}^k_{(ir, i\theta)}
  \mathcal{V}^k_{(ir+1, i\theta+1)}}{\mathcal{V}^k} \,, 
\end{eqnarray}
where the sum is considered over all the $k$ particles located into the
volume $\mathcal{V}$, and $v^j = \gamma^{ij} u_i / \Gamma$ is the
particle's three-velocity measured by FIDOs. In a similar way, one can
compute the charge and current densities on the other Yee grid nodes.

\subsection{Divergence cleaning}

The Yee algorithm guarantees that the divergence-free constraint on the
magnetic field is preserved exactly in time, i.e., $\partial_t (\nabla
\cdot \boldsymbol{B}) = 0$, provided it is satisfied initially. In
contrast, Gauss law for the electric displacement field, $\nabla \cdot
\boldsymbol{D} = 4\pi\rho$, is not automatically preserved in the
presence of charged particles, unless the current deposition scheme
satisfies the discrete charge continuity equation exactly. As a
consequence, numerical errors in $\nabla \cdot \boldsymbol{D}$ may
accumulate over time and require explicit correction.

To enforce Gauss's law, we periodically perform divergence cleaning by
solving the Poisson equation
\begin{equation}
\dfrac{1}{\sqrt{\gamma}} \partial_k \left( \sqrt{\gamma} \gamma^{kl}
\partial_l \phi \right) = \dfrac{1}{\sqrt{\gamma}} \partial_j \big(
\sqrt{\gamma} D^j \big) - 4 \pi \rho\,, 
\label{phi}
\end{equation}
and correcting the electric field according to
$D_{\mathrm{new}}^i = D_{\mathrm{old}}^i - \gamma^{ij} \partial_j \phi$.
In \texttt{FPIC}, Eq.~(\ref{phi}) is solved using an iterative Jacobi method.
We usually update the correct electric field every 25 timesteps, solving 
Eq.~(\ref{phi}) with 500 iterations, but these values can change depending 
the physical setup employed.

\subsection{Boundary conditions}

\begin{figure}
\centering
\includegraphics[width=0.8\columnwidth]{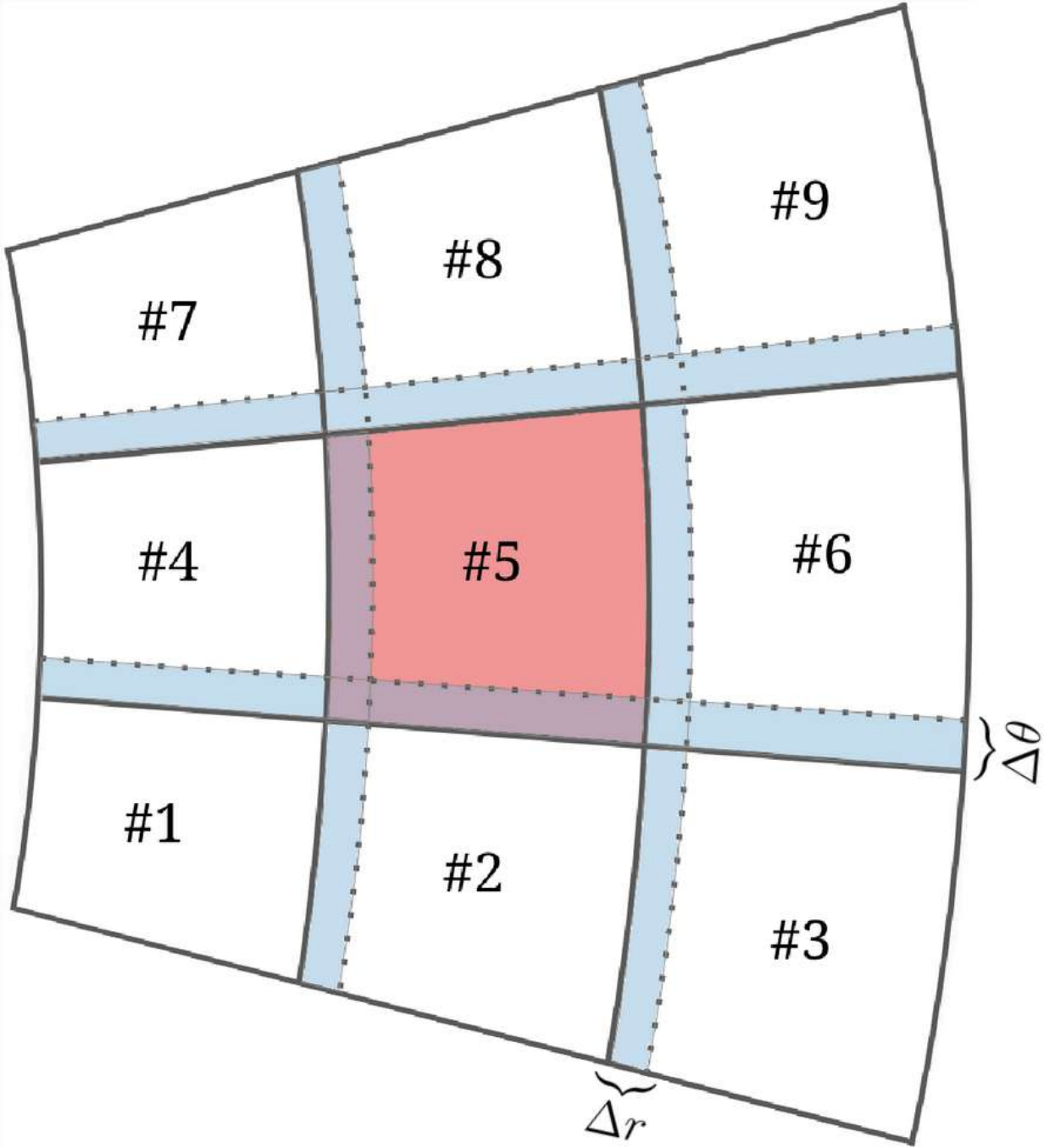}
\caption{MPI-domain decomposition scheme employed in \texttt{FPIC}. The
  diagram shows a portion of the computational domain with 9 CPUs, where
  solid black lines delimit the \textit{physical} sub-domains assigned to
  each CPU. Each \textit{computational} sub-domain extends beyond the
  physical one by $\Delta r$ (radial direction) and $\Delta \theta$
  (angular direction) to allow MPI communications, and CPUs exchange
  information with the neighbours across the light-blue shaded regions.}
\label{proc} 
\end{figure}

Thanks to the properties of the Kerr-Schild coordinates, the inner radial
boundary is set to be within the event horizon, which lies in a region
causally disconnected from the rest of the domain, and here we enforce a
zero-gradient constraint on the dynamical variables $\partial_r
{\boldsymbol{D}} = \partial_r {\boldsymbol{B}} = 0$. At the outer
boundary of the radial domain, we aim to mimic \text{open} boundaries, 
namely perfectly absorbing boundaries with no reflection of waves and
particles, thus isolating the system from the external environment. In
practice, we add an artificial dissipative term to Maxwell's equations to
exponentially damp the fields in a spherical shell located at the outer
regions of the box, according to~\citep{Cerutti2015}
\begin{eqnarray}
\partial_t {\boldsymbol{B}} &=& - \nabla \times {\boldsymbol{E}} -
\chi(r) {\boldsymbol{B}}\,, 
\label{abs0}
\\
\partial_t {\boldsymbol{D}} &=& \nabla \times {\boldsymbol{H}} - 4 \pi
        {\boldsymbol{J}} - \chi(r) {\boldsymbol{D}}\,, 
\label{abs1}
\end{eqnarray}
where $\chi(r)$ is the conductivity of the absorbing region and reads
\begin{equation}
\chi(r) = \chi_0 \left(
\dfrac{r-r_{\text{abs}}}{r_{\text{max}}-r_{\text{abs}}}
\right)^{\lambda}\,.
\label{abs2}
\end{equation}
Typical values are $r_{\text{abs}}=0.9 \, r_{\text{max}}, \chi_0=10^4$, 
and $\lambda=5$, but they may vary slightly across different setups.

The $\theta$-domain can assume two geometries: the \textit{full}-$\theta$
domain, covering the range $0 \leq \theta \leq 2 \pi$, and the
\textit{half}-$\theta$ domain, extending from the south pole ($\theta_0 =
0$) to the north pole ($\theta_{\text{max}} = \pi$). The field and
particle boundary conditions change accordingly: In the
\textit{full}-$\theta$ configuration, we simply apply periodic boundary
conditions at $\theta_{\text{max}}= 2\pi$; for the \textit{half}-$\theta$
configuration we apply the axial symmetry to all Yee field components
located on the boundaries $\theta_0$ and $\theta_{\text{max}}$, \ie
$\partial_r D^r = D^\varphi = B^\varphi=0$. To avoid numerical issues
when $\theta$ approaches the poles, we set $\theta_0 = \epsilon$ and
$\theta_{\text{max}}=\pi - \epsilon$, typically with $\epsilon =
\pi/200$. Similarly, in the \textit{full}-$\theta$ geometry, we set
$\theta_0 = \epsilon$ and $\theta_{\text{max}}= 2\pi + \epsilon$, with
$\epsilon= \pi/(2 N_\theta)$, ensuring that the poles lie midway between
two Yee-grid points.

Particles falling into the black hole ($r < r_\text{h}$) or reaching the
outer boundary ($r > r_{\text{max}}$) are removed from the
simulation. They are reflected upon passing across the poles for the
\textit{half}-$\theta$ domain configuration (\ie with $u_\theta$ changing
sign at the poles), while they follow periodic boundary conditions for
the \textit{full}-$\theta$ configuration.

\begin{figure*}
\includegraphics[width=1.0\hsize]{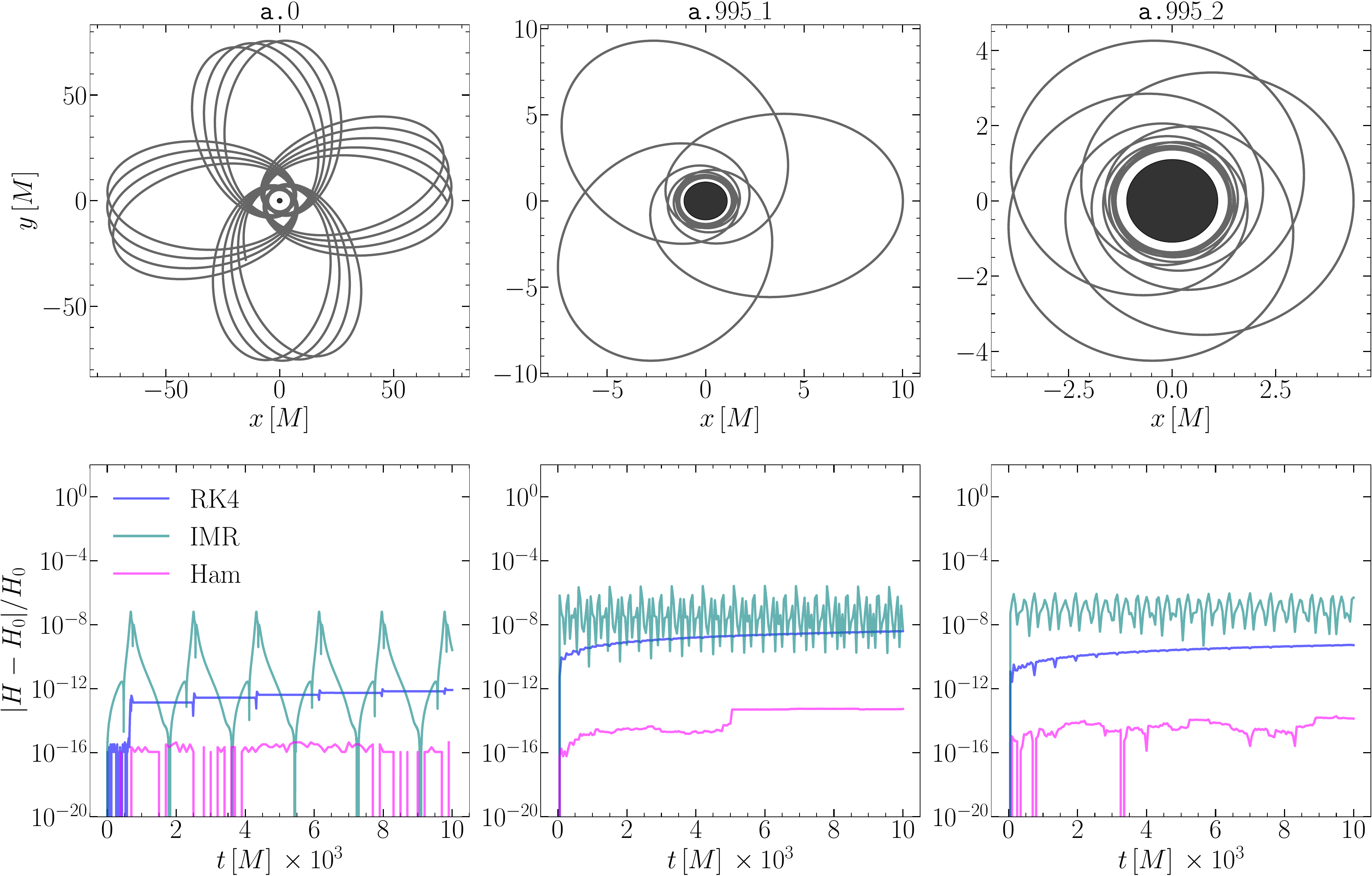}
\caption{Top row: orbits in the $(r,\varphi)$ plane for neutral particles
  around black holes. The black disk at the center of each panel
  represents the horizon; Bottom row: deviation of energy from its
  original value, $|H-H_0|/H_0$, vs time for each case. From left to
  right: a few precessing orbits for the ``four-leaf'' orbit with
  $E=0.987649$ and $L=3.9$ around a Schwarzschild black hole; orbits
  around Kerr black hole with $a_*=0.995$, and $\{E, L\}=\{0 .916235,
  2\}, \{0 .841423, 1.82\}$, respectively (see the Tab.~\ref{tab:params}
  for details). We report the energy errors for the RK4 (blue), the IMR
  (teal), and the Hamiltonian (magenta) schemes, up to $t=10^4 \, M.$}
\label{part0}
\end{figure*}

\subsection{Parallelisation}

\texttt{FPIC} implements a grid-based domain decomposition applied
through the MPI directives, which ensures the communication among the
different processors. The global simulation box is described by $N_r
\times N_\theta$ grid points, divided into a number of CPUs given by
$N^\text{cpu} = N^{\text{cpu}}_r \times N^{\text{cpu}}_\theta$, where the
latter refer to the CPUs along the $r-$ and $\theta-$ directions,
respectively. In a spherical 2D decomposition, the bulk domains have 8
neighbours (see Fig.~\ref{proc} for a schematic representation). Each of
the $N$ CPUs is assigned to a portion of the entire domain, evolving the
fields and pushing the particles that propagate within its proper
sub-domain, given by $(N_r/N^\text{cpu}_r+1) \times
(N_\theta/N^\text{cpu}_\theta+1)$. The $``+1''$ refers to the so-called
\textit{ghost cells} (the blue shaded regions in Fig.~\ref{proc}), which
are needed for communication among neighbours. This means that the
domains overlap in such a way that they share common cells, so that
neighbour CPUs can exchange information such as which particles have left
or entered their domain, and what the updated values of the fields are at
the boundaries of the domain. The latter is used, for instance, when
computing derivatives at the edges of each sub-domain.

\section{Code Validation}
\label{Sec3}

In this Section we aim to test the \texttt{FPIC} code through a series of
numerical testbeds. We start by validating the particle pushers by
integrating the trajectories of neutral particles under the influence of
the gravitational field of the black hole only. Then we evolve the
dynamics of charged test-particles when immersed in a stationary Wald
magnetic field, monitoring the conservation of the Hamiltonian
energy. Finally, we test the Maxwell equations solver both in
electrovacuum and in a plasma-filled scenario, by using a Wald and a
split-monopole configuration.

\subsection{Test Particle Trajectories}
\label{testpart}

\subsubsection{Neutral particles}
\label{sec_neutralpart}

\begin{figure*}[t]
\centering
\includegraphics[width=1.0\hsize]{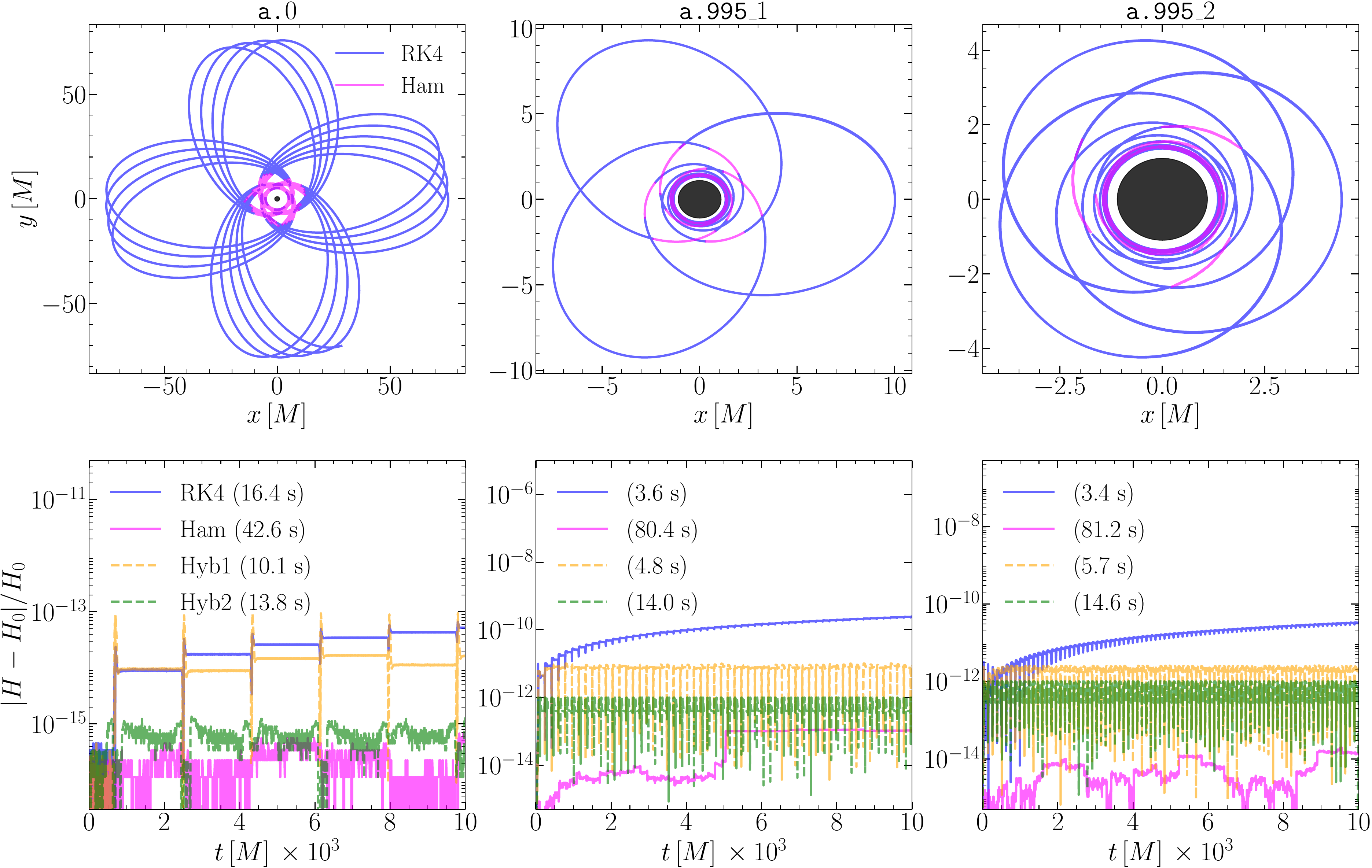}
\caption{Same trajectories for neutral particles as reported in
  Fig.~\ref{part0}, computed using our hybrid approaches. Top row:
  particle trajectories colour-coded according to the integrator adopted
  for the ``Hyb2'' scheme (see text for details and
  Tab.~\ref{tab:params_hyb} for the numerical parameters). The RK4
  integrator (blue) is employed when the particle is far from the black
  hole, while the Hamiltonian scheme (magenta) is primarily used as the
  particle enters stronger gravitational-field regions. Bottom row:
  energy violations obtained using fixed integrator schemes (solid lines)
  and our hybrid approaches (dashed lines). The total runtime, in
  seconds, is also reported for each case.}
\label{hyb}
\end{figure*}

We integrate the paths of particles around black holes to test particle
solvers implemented in \texttt{FPIC}. As a first testbed, we consider
only the geodesic solver, in the absence of any electromagnetic field. We
have selected three particular paths for neutral particles (\ie $q=0$ in
Eq.~(\ref{O2})) both in Schwarzschild and Kerr metrics~\citep{Levin2008,
  Bacchini2018, Chen2025} (for details of physical and numerical
parameters, see the Tab.~\ref{tab:params}). The particle's motion lies in
the equatorial plane, at $\theta_0 = \pi/2$, with vanishing
$u_\theta$. Among the three orbits, the first one is for a Schwarzschild
black hole, while the others refer to a Kerr black hole with spin $a_* =
0.995$. Correctly reproducing these trajectories is a first but important
testbed for the particle pushers implemented in \texttt{FPIC}, being a
test of long-time stability thanks to the periodicity of orbits. For
these tests only the term $\mathcal{A}_\text{g}$ acts on the particle,
and we compute the metric components and their derivatives analytically,
directly on the particle position. Since no interpolations are needed,
the number of mesh points do not play any role on the accuracy of the
simulations. The timestep is fixed to $\Delta t = 10^{-1}$ for all the
three orbits.

In the top row of Fig.~\ref{part0} we report the particle trajectories
which are produced with \texttt{FPIC}, and integrated up to $t = 10^4
M$. Note that the trajectory around the Schwarzschild black hole slowly
precesses and close up on itself after approximately 1000 cycles (we
report only a few circles of the trajectory), while for the orbits around
Kerr black holes we report the full trajectories up to $t=10^4 \, M$.
The panels in the bottom row show the evolution of $|H(t)-H_0|/H_0$
accordingly, where $H = -u_0$ is the conserved energy and $H_0$ is the
initial energy. The different colours refer to different integrators,
highlighting that the Hamiltonian scheme is the most accurate among the
others. The RK4 and the IMR method show higher violation of energy, but
the error is overall small and stable up to long timescales. All the
schemes show a good conservation of the energy over the entire duration
of integration for all the particle trajectories, and are consistent with
other approaches in the literature~\citep{Levin2008, Bacchini2018,
  Chen2025}. Even if the RK4 scheme is non-symplectic, it is the faster
one among the 3 integrators, and we measured
$t_{\text{IMR}}/t_{\text{RK4}} \sim 2.15$ and
$t_{\text{Ham}}/t_{\text{RK4}} \sim 41.2$, where $t_\#$ refer to the time
needed to reach $t=10^4\, M$ for each scheme.

\subsubsection{Hybrid integration scheme} 
\label{sec_hyb}

\begin{figure*}
  \includegraphics[width=1.0\hsize]{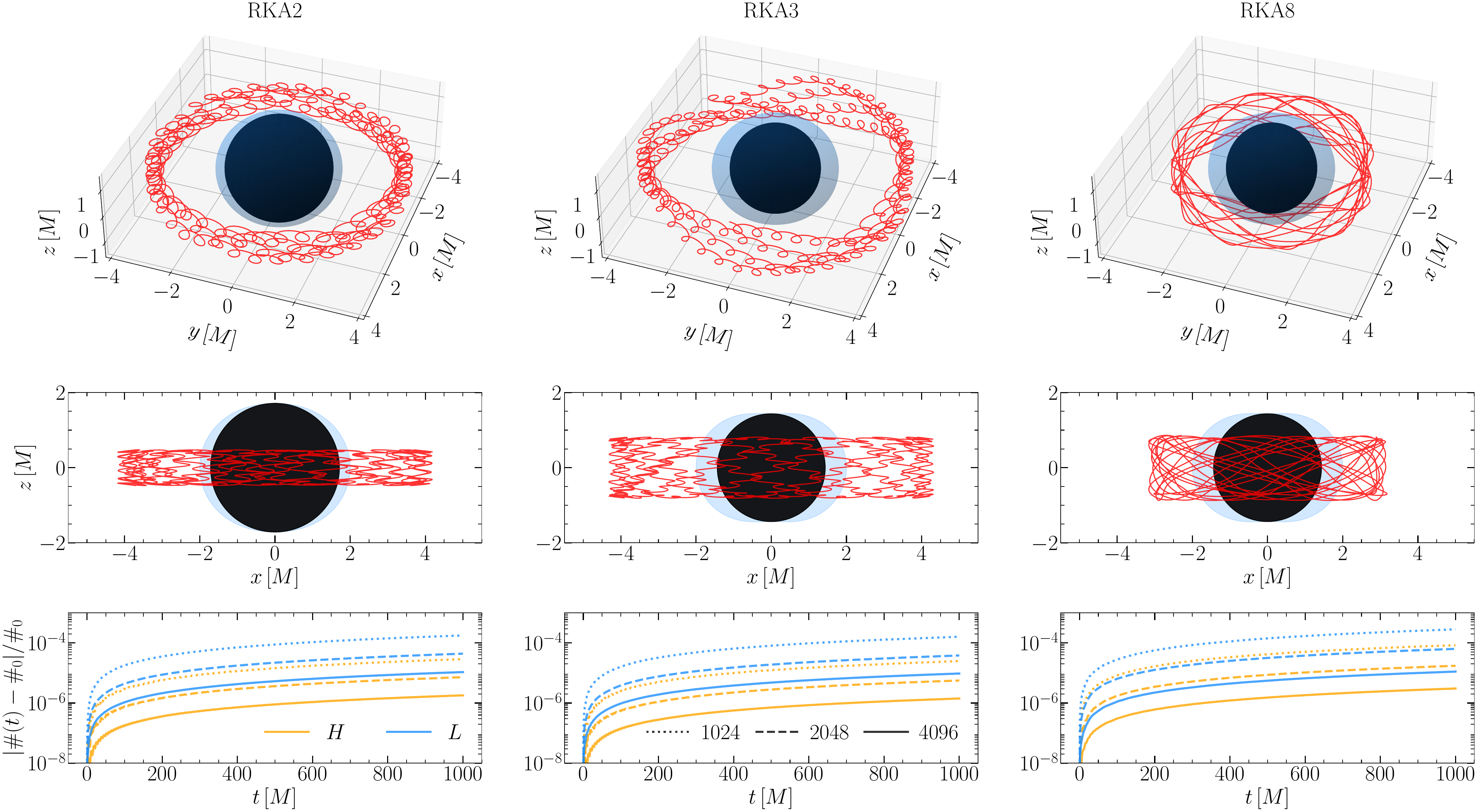}
  \caption{Top: 3D charged particle trajectories around Kerr black holes,
    for different cases; Central: projection on the $(x,z)$ plane, for
    the same trajectories; Bottom: conservation of $H$ (yellow) and $L$
    (blue) for different spatial resolutions $N=N_r=N_\theta$, according
    to the legend. The horizon is reported as a black sphere, and the
    ergosphere as a blue shaded area. The physical and numerical
    parameters are reported in Tab.~\ref{tab:params}. }
\label{fig3d}
\end{figure*}

As we have seen, the Hamiltonian scheme is very computationally demanding, 
but it is able to retain extremely low values of errors.
Conversely, the RK4 integrator is much faster, though at the price of a larger
error violation. In particular, the bottom-left panel of Fig.~\ref{part0}
shows an interesting behavior in the RK4 error evolution: the violation remains
almost constant in time, except at a few specific moments, 
\ie when the particle approaches the \textit{periastron}, and the energy error
exhibits a sudden ``jump'' before returning to a constant level.
This evolution - in a sort of Heaviside-like fashion - suggests that the RK4
scheme performs reasonably well for most of the simulation, except when the
particle experiences stronger gravitational forces where a more accurate
integrator is required. In light of this, we propose a hybrid scheme that 
can self-consistently switch between the RK4 and the Hamiltonian integrator.

The key idea behind the hybrid scheme is simple yet efficient: at each
timestep $n$, we monitor the violation of the Hamiltonian
$\widetilde{H}^{n+1}:= |H(t^{n+1})-H_0|/H_0$ and select the most suitable
integrator accordingly. We first update the particle positions and
momenta with the RK4 method to obtain the Hamiltonian energy at timestep
$n+1$. If a prescribed condition is satisfied, the updated particle
position and velocity are accepted, otherwise, the step is rejected and
the Hamiltonian integrator is employed instead. We consider two
alternative criteria for this condition: the first, constrains the time
derivative $\partial_t \widetilde{H}^{n+1}<(\partial_t
\widetilde{H})_{\text{max}}$, while the second, directly constrains the
value $\widetilde{H}^{n+1}<\widetilde{H}_{\text{max}}$. We will refer to
these two approaches as ``Hyb1'' and ``Hyb2'', respectively (See
Tab.~\ref{tab:params_hyb} for details on numerical parameters.)

In Fig.~\ref{hyb} we show the results for our hybrid approach, for the
same trajectories shown in Fig.~\ref{part0}. In the first row  
we report the trajectory for each case, colour-coded
according to the integrator employed when using the ``Hyb2''
scheme. Notice that the RK4 scheme is fully employed when the particle's
trajectory is far from the black hole, while it switches to the
Hamiltonian scheme as the particle approaches it. In the bottom row we
present a comparison of the Hamiltonian violation for different
integration schemes, according to the legend, where we also report the
total runtime (in seconds), when fixing all the other numerical
parameters. As a reference and for all the three cases, the energy
violation for the RK4 and Hamiltonian schemes are reported with solid
lines in blue and magenta, respectively. Two additional hybrid
simulations are shown with dashed lines, taking advantage of the
stability of the Hamiltonian integrator at larger timesteps and using
adaptive timesteps, reporting results for ``Hyb1'' (orange lines), and
``Hyb2'' (green lines).

Interestingly, the hybrid approach is able to speed up the simulation
while retaining a smaller error violation, for example, comparing
``Hyb1'' (orange line) and ``RK4'' (blue line) for the case
$\texttt{a.}0$ in the left panels of Fig.~\ref{hyb}. Alternatively, it is
possible to achieve even smaller error violations at a slightly higher
computational cost, as shown for the ``Hyb2'' case (green line). In this
case, the ratio between computational cost and accuracy improves
significantly, with relative errors remaining at the level of $\lesssim
10^{-15}$, and a substantial speedup compared to the RK4 scheme, as well
as the Hamiltonian integrator (magenta line).

Similar strategies can also be applied to the spinning cases
$\texttt{a.995}\_1$ and $\texttt{a.995}\_2$ (shown in the middle and
right columns of the same Figure). Here, the performance gain is reduced
because the trajectory remains closer to the black hole compared to the
Schwarzschild case, but the hybrid approaches still represent a valid
compromise between speedup and accuracy.

\subsubsection{Charged particles}
\label{sec_chargedpart}

Next step is to combine the effects of gravitational and electromagnetic
fields which is essential for realistic PIC simulations. Following the
approach of~\citet{Bacchini2019, Chen2025}, we evaluate the paths of
charged particles embedded in a vacuum Wald field solution, with initial
four-potential given by~\citep{Komissarov2004b}:
\begin{equation}
A_\mu= \dfrac{B_0}{2} \left( g_{0 \varphi} + 2 a_* g_{00}, g_{r \varphi} 
+ 2 a_* g_{0r}, 0, g_{\varphi \varphi} + 2 a_* g_{0 \varphi} \right)\,.
\label{Amu}
\end{equation}
From the above expression, the $\boldsymbol{E}$ and $\boldsymbol{B}$ field can be obtained by:
\begin{eqnarray}
B^i &=& \dfrac{1}{\sqrt{\gamma}} \eta^{ijk} \partial_j A_k\,, 
\label{wald0}
\\
E_i &=& -\partial_i A_0 - \partial_0 A_i\,, 
\label{wald1_0}
\end{eqnarray}
and reported in detail in the Appendix~\ref{ICwald}. Following the
notation of~\citet{Chen2025}, we have selected the non-chaotic orbits
labelled as RKA2, RKA3, and RKA8. Both the metrics are described by Kerr
black holes, with dimensionless spin parameter $a_*=0.7$ for RKA2, and
$a_*=0.9$ for RKA3 and RKA8. Even if the gravitational and
electromagnetic fields are stationary and analytically known, here we aim
to test the interpolation scheme in \texttt{FPIC}, as done in typical
kinetic simulations. We start with $N_r = N_\theta = N=1024$ collocation
points, with a radial domain covering $0.99 \, r_h \leq r \leq 6\, M$ and
a polar one $1/200 \leq \theta/\pi \leq 199/200$. For all the cases we
use a fixed timestep $\Delta t=10^{-3}$, and we set $q=m=1$ for the
particle's charge and mass. We use the RK4 scheme integrating the
trajectories up to $t=10^3 \, M$, which is greater than the typical
timescales of GRPIC simulations.

\begin{figure*}    
\centering
\includegraphics[width=1.0\hsize]{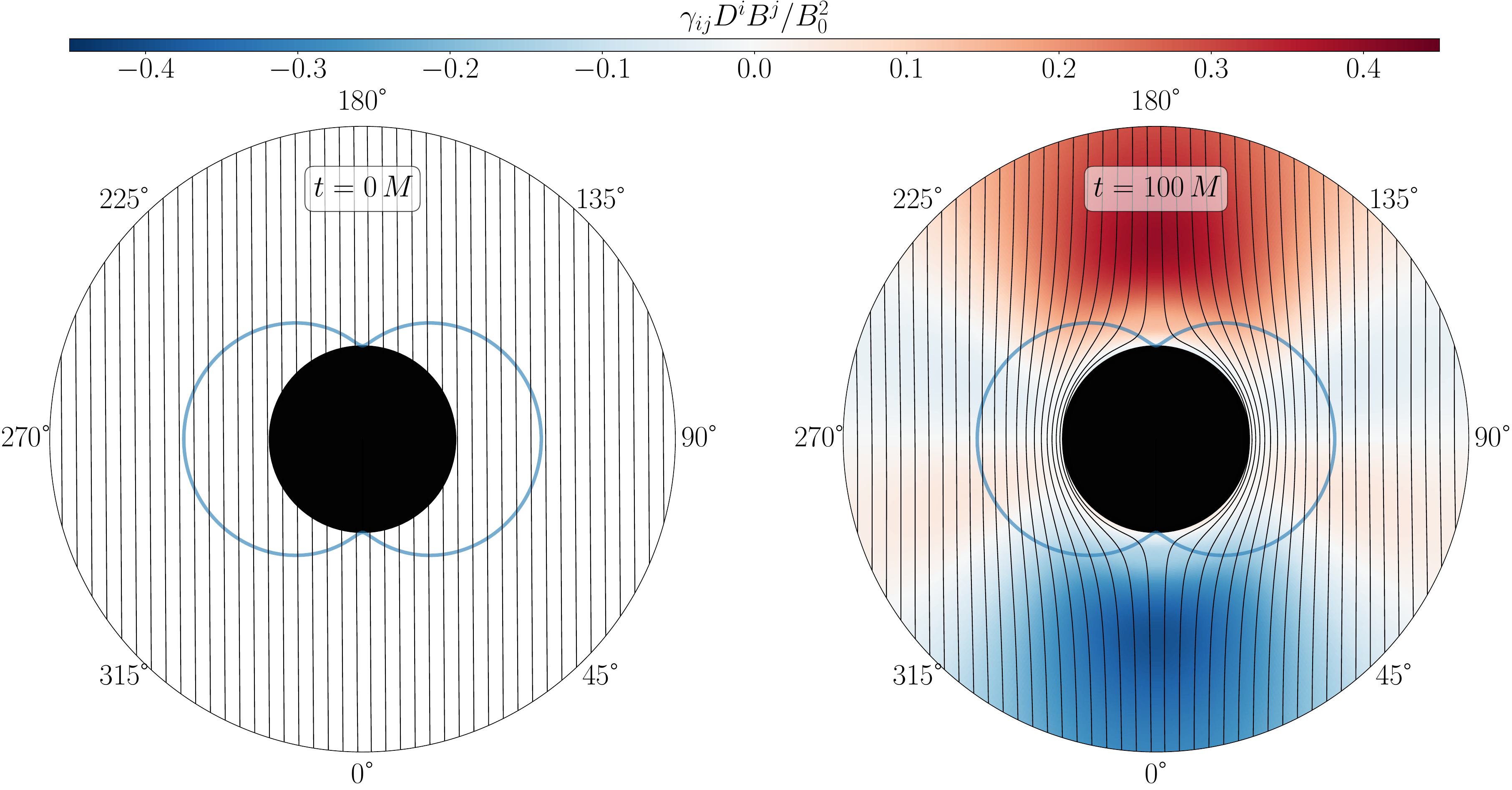}
\caption{Vacuum Wald solution for a rotating black hole with spin $a_* =
  0.999$. Left: Initial state, which is the Wald solution for a
  non-rotating black hole; Right: Steady state at $t=100 \, M$. The
  colormap report the parallel electric field $D_i B^i/B_0^2$. Black
  solid contours show magnetic-field lines (being the isocontours of the
  potential vector $A_\varphi$). The solid black sphere shows the horizon
  of the black hole and the light-blue curve shows the ergosphere. In
  the steady state (right panel), magnetic-field lines are expelled from
  the horizon.}
\label{fig:wald_vuoto}
\end{figure*}

A 3D visualization of the integrated charged trajectories is shown in the
top row of Fig.~\ref{fig3d} for the different cases, while in the central
row we report a polar projection, in the $(x,z)$ plane. Note that for
each case, we report the horizon as a black central sphere, and the
ergosphere as a light-blue shaded area. We also computed the energy
conservation properties of the particle integrator in \texttt{FPIC} for
charged particle trajectories, as the Hamiltonian $H$ and the angular
momentum $L$. In the presence of the electromagnetic field, the vector
potential must be evaluated and added in the expression for the conserved
energy, according to:
\begin{equation}
H = -u_0 - \dfrac{q}{m} A_0\,, ~~~A_0 = B_0 \dfrac{a_* r (1+\cos^2
  \theta)}{r^2 + a_*^2 \cos^2 \theta}\,.
\end{equation}
Similarly, the conserved angular momentum becomes:
\begin{equation}
L = u_\varphi + \dfrac{q}{m} A_\varphi\,, ~~~A_\varphi = \dfrac{\sin^2
  \theta}{2} \left[ r^2 + a_*^2 - \dfrac{2 a_*^2 r (1+cos^2 \theta)}{r^2
    + a_*^2 \cos^2 \theta} \right]\,.
\end{equation}
In the bottom row of Fig.~\ref{fig3d} we report both the conservation of
energy (yellow lines) and angular momentum (light-blue lines) for each case,
when a spatial resolutions of $N=1024, 2048$ and $4096$ is used. The
errors clearly decrease with increasing the number of mesh points,
indicating that the conservation of $H$ and $L$ is essentially dominated
by the interpolation of field values from the grid points to the particle
positions. As a confirmation, we measured identical errors, when halving
or doubling the timestep, with a fixed number of grid points, or when
using the more demanding IMR scheme (not shown here). In light of this,
we found the RK4 integrator to be the best compromise, as it is the least
computationally demanding, and using a higher-order particle integrator
provides limited benefit at a fixed spatial resolution. Finally, it is
worth mentioning that typical values for large-scale GRPIC simulations
are $N \gtrsim 1024-4196$, $\Delta t \lesssim 10^{-4} \, M$, and
$t_{\text{final}} \sim 100-200 \, M$, making these errors adequate for
our purposes.

\subsection{Vacuum Wald Solution} 
\label{waldvac}

To test the correctness of our Maxwell equation solver, we simulate a
uniform magnetic field in electrovacuum. This choice is astrophysically
motivated from currents located in a remote accretion disk naturally
producing a large-scale magnetic field which is almost uniform at the
black hole scale. The steady state solution assumes that the energy
density of the field is small enough so that it does not significantly
change the spacetime metric. In the case of an external field aligned
with the spin axis of the black hole, the solution was found by
\citet{Wald:74bh}, and was generalized to arbitrary inclinations by
\citet{Bicak:1985}.

\begin{figure*}    
\centering
\includegraphics[width=1.0\hsize]{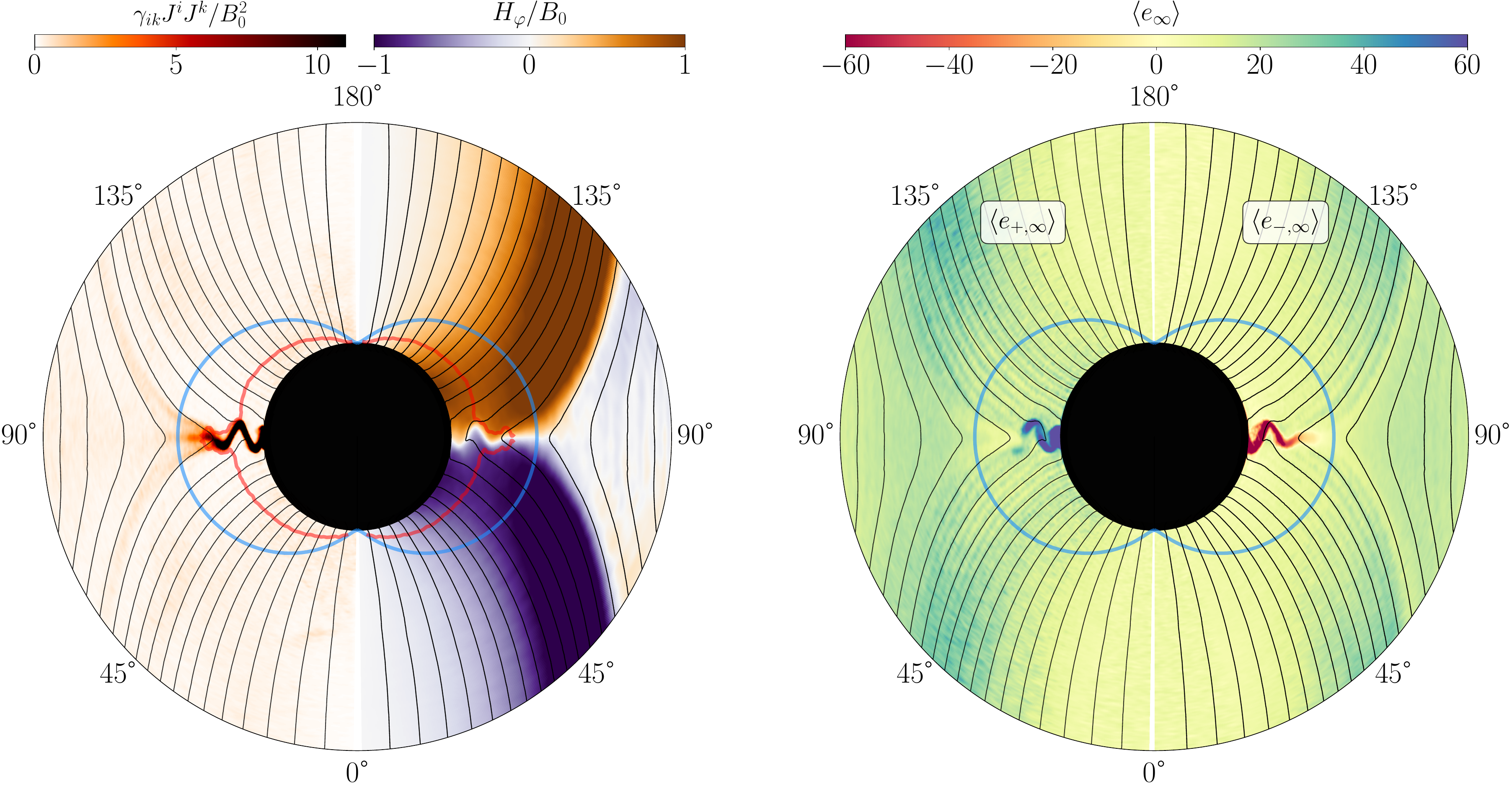}
\caption{Plasma-filled Wald solution for black hole spin $a_* = 0.999$, 
  at $t=40\, M$. Left panel: square of the total current density
  ${{J}}^2/B_0^2$ (left), and toroidal magnetic field
  $H_\varphi/B_0$ (right). Shown in red there is the inner light
  surface, always located inside the ergosphere (reported with light-blue
  line). Right panel: averaged energy at infinity $e_{\infty}=-u_0$ for
  positrons (left) and electrons (right). Negative-energy electrons are
  present inside the current sheet.}
\label{fig:wald_full}
\end{figure*}

We initialise the axisymmetric computational domain starting from within
the event horizon, thus in a region causally disconnected from the rest
of the domain, with a radial extent of $0.99 \, r_h \leq r \leq 20 M$ and
a polar one covering the full angle of $0 \leq \theta \leq 2\pi$, via
$N_r \times N_\theta = 4196 \times 1024$ cells. An additional outer layer
with absorbing boundary conditions for the electromagnetic fields is
applied at $r=r_{\text{abs}}= 0.9 \, r_{\text{max}}$, according to
Eqs.~(\ref{abs0}) and (\ref{abs1}). We immerse a rotating black hole with
dimensionless spin parameter $a_* = 0.999$ into a uniform Wald-solution
field with strength $B_0=10^3$. Even if the spacetime is described by a
highly rotating Kerr metric, the initial Wald field is the solution for a
Schwarzschild black hole (\ie we set $a_*=0$ in Eq.~\eqref{Amu}), and let
the system evolve. The electric field measured on the grid is initially
$E_i=0$, while the electric field measured by FIDOs is nonzero and reads
${\boldsymbol{D}}=-{\boldsymbol{\beta}} \times {\boldsymbol{B}} / \alpha$, 
according to Eq.~(\ref{Ei}). In the left panel of
Fig.~\ref{fig:wald_vuoto} we report the initial state of the simulation, 
with the parallel electric field $D_i B^i$ vanishing everywhere, while
vertical lines represent the magnetic field.

Since the initial magnetic configuration is not solution for a spinning
black hole, the field naturally evolve and establish a new steady state
which agrees with the spinning Kerr metric, converging towards the Wald
configuration. Right panel of Fig.~\ref{fig:wald_vuoto} shows the steady
state obtained at $t=100 \, M$. The solution is characterized by the
expulsion of the magnetic-field lines away from the horizon, phenomenon
often called the \textit{Meissner effect}. The parallel electric field
$D_i B^i/B_0^2$ is no more vanishing as in the initial state, because the
rotation of the black hole induces a strong unscreened electric field
parallel to the magnetic field, as reported in the right panel of Fig.
\ref{fig:wald_vuoto}. The electric field is distributed according to a
quadrupolar-like angular distribution, with maxima located around the
polar regions and reaching values of the order of $D_iB^i \lesssim
B_0$. The occurrence of gravitationally induced parallel electric fields
$D_iB^i$ is another major feature of the Wald solution, and it is capable
of accelerating particles and driving currents.

\subsection{Plasma-filled Wald Solution}
\label{waldfull}

The natural extension is to simulate a plasma-filled magnetosphere of a
spinning black hole that is threaded by a uniform magnetic field
\citep{Parfrey2019, Chen2025}. We initialise the electromagnetic fields
with the exact Wald solution given by Eqs.~(\ref{wald1})--(\ref{wald6}), 
and setting $a_*=0.999$. The computational domain covers a radial extent
of $0.99 \, r_h \leq r \leq 20 M$ and a polar one of $1/200 \leq
\theta/\pi \leq 199/200$, via $N_r \times N_\theta = 4196 \times 1024$
cells. The magnetosphere is initially in electrovacuum, and we inject
pair plasma into the spherical volume with coordinate radius $r_h \leq r
\leq 6\, M$. In each computational cell, an electron-positron pair is
added provided that $n < \mathcal{M} \, n_{\text{GJ}}$, where
$\mathcal{M}$ is the so-called ``multiplicity'', $n=n_{e^-}+n_{e^+}$ is
the total plasma density, $n_{e^-}, n_{e^+}$ are the electron and positron
number density, respectively, $n_{\text{GJ}} = -{\boldsymbol{\Omega}}
\cdot {\boldsymbol{B}}/(4 \pi e)$ is the Goldreich-Julian number density
\citep{Goldreich:1969}, which measures the minimum number density
required to screen the longitudinal components of the electric field, and
$\boldsymbol{\Omega}$ is the angular velocity of the event horizon. We
perform the injection of new particles every $\Delta t_{\text{inj}}=
0.01\, M$ conferring an effective FIDO-measured density of $n_0 = 5 \, 
n_{\text{GJ}}$, and only for those cells for which the multiplicity
condition $\mathcal{M}<3$ is satisfied. We randomly draw the FIDO-frame
particle velocities from a relativistic Maxwell--J\"uttner of
dimensionless temperature $\Theta := k_\text{B} T/m_e = 0.5$, with
$k_\text{B}$ being the Boltzmann constant.
 
The initial magnetic-field strength $B_0$ is specified in terms of the
dimensionless field $\tilde{B}_0 := r_{g} /r_{_{\rm L}} = GM\, eB_0 /(m_e
c^4)$ where $r_{g}:=GM/c^2$ is the gravitational radius, $r_{_{\rm L}} :=
m_e c^2/(eB_0)$ is the Larmor radius. Following~\citet{Parfrey2019}, we
set $\tilde{B}_0 = 10^3$, which is equivalent to a magnetic field of
$\simeq 1\, {\rm G}$ for a black hole of mass $M\simeq 10\, M_{\odot}$ (or
of $\simeq 10^{-5}\, {\rm G}$ for a black hole with the mass of Sgr~A*, 
\ie $M\simeq 10^{6}\, M_{\odot}$). With our choice of magnetic-field
strength and spin, the ``cold'' plasma magnetisation is
$\sigma_0=B_0^2/(4 \pi n_{\text{GJ}}m_e) \simeq 2000$.

\begin{figure*}[t]
\centering
\includegraphics[width=1.0\hsize]{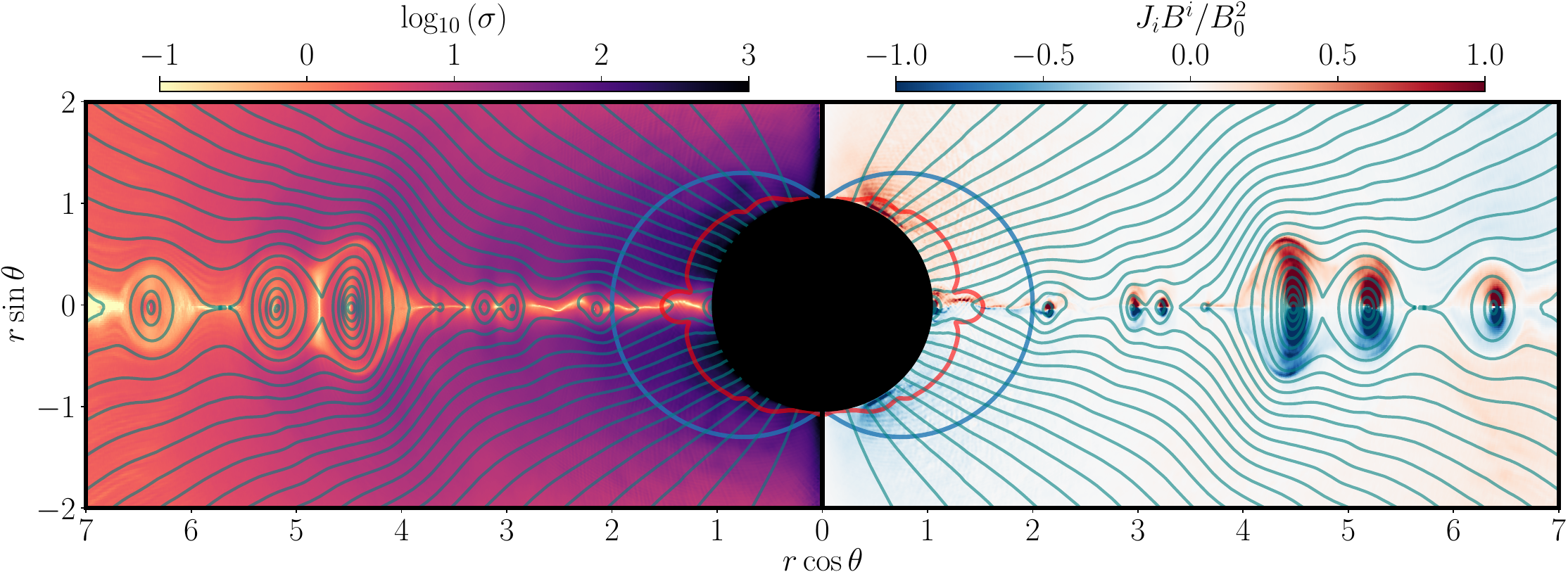}
\caption{Overview of a representative simulation with spin parameter
  $a_*=0.999$, at time $\bar{t} = 14\, M$. \textit{Left panel:} ``cold''
  plasma magnetisation $\sigma :=\gamma_{ij}B^i B^j/(4 \pi n
  m_e)$. \textit{Right panel:} alignment between the current density and
  the magnetic field, $\gamma_{ij} J^i B^j / B_0^2$. Also shown are the
  ergosphere (light-blue), the inner light surfaces (red), and magnetic
  field lines (teal). The current sheet is located at the equatorial
  plane where magnetic reconnection takes place and a plasmoid-chain
  forms. Note that two plasmoids are going to merge at $r \sim 5\, M$. }
\label{split} 
\end{figure*} 

In contrast to the vacuum case where the magnetic-field lines were
expelled out from the horizon, now the steady state is composed of a
series of field lines threading the ergosphere, dragged and twisted by
the rotation of space time, with a significant fraction passing through
the black hole horizon. An equatorial current sheet quickly forms at the
horizon, where the plasma falls along the field lines toward the black
hole. At $t \sim 40 \, M$ the current sheet extends to the ergosphere
boundary, being disrupted by the drift-kink instability (left panel of
Fig.~\ref{fig:wald_full}, left side, where we report $J^2/B_0^2$). When
the steady state is reached, the toroidal magnetic field $H_\varphi$ is
large in the jet ($H_\varphi \sim B_0$), and very small outside it, as
reported in the left panel of Fig.~\ref{fig:wald_full} (right side) for
$t = 40 \, M$.  Reported in light-blue is the ergosphere, and 
in red the inner light surface
(always located inside the ergoregion), where the function
$\mathcal{L}_{\rm ls} (\Omega_{f}, r, \theta) := g_{\phi\phi}
\Omega_{f}^2 + 2 g_{t \phi}\Omega_{f} + g_{t t}=
0$~\citep{Komissarov2004b}, being
$\Omega_{f}=-E_\theta/(\sqrt{\gamma}B^r)$ the angular velocity, and
$g_{\mu \nu}$ is the background four-metric.

Rapidly rotating black holes naturally show the presence of particles
with negative energy at infinity, \ie $e_{\infty} := -u_0 < 0$, where
$u^\mu$ is the particle's four-velocity~\citep{Penrose69, Penna2015, 
  Hirotani2021, ElMellah2021}). Under these conditions, a Penrose process
is possible, whereby energy can be extracted from a Kerr black hole when
the negative-energy particles cross the event
horizon~\citep{Penrose69}. We report the cell-averaged energy at infinity
$\langle e_{\infty} \rangle$ in the right panel of
Fig.~\ref{fig:wald_full} for positrons (left) and electrons (right).
Electrons with $\langle e_{\infty} \rangle<0$ are ubiquitous in the
current sheet, eventually falling toward the horizon and extracting
rotational energy, according to the Penrose process.

\subsection{The Split Monopole}
\label{SM}

For the split monopole case we initialise the radial component of the
FIDO magnetic field as $B^r := \partial_\theta\Psi/\sqrt{\gamma}$, where
$\Psi(r, \theta)$ is the gauge-invariant magnetic-flux function $\Psi(r,
\theta) = 4M^2B_0 \zeta(\theta)(1-\cos\theta)$ (see the Appendix
\ref{ICsplit} for details). The factor $\zeta(\theta)$ is introduced to
model the equatorial current-sheet necessary to sustain the field
discontinuity~\citep{Gralla2014, Meringolo2025a}. In the presence of
spacetime rotation, the initial split-monopole evolves towards a
stationary electrovacuum configuration that preserves a monopolar
topology in the polar plane and an equatorial current-sheet, while
exhibiting gravitationally-induced toroidal magnetic fields, as well as
electric fields along the magnetic-field lines. These gravitationally
induced electromagnetic fields are similar to what observed in the Wald
solution~\citep{Wald:74bh} as considered, for instance, in
\citet{Moesta:2009, Alic:2012, Parfrey2019}.

After the vacuum electromagnetic fields settle to a stationary
configuration at $t=t_{\rm inj} = 50\, M$, a $e^-/e^+$-pair plasma is
injected in the spherical volume with coordinate radius $r_{h} < r <
15\, M$. In each computational cell, an electron-positron pair is added
for all subsequent times $\bar{t} := t - t_{\rm inj}>0$, provided that
the condition $n < \mathcal{M}\, n_{_{\rm GJ}}$ is satisfied, and we set
the multiplicity to $\mathcal{M}=10$ so as to have an accurate
representation of the black hole magnetosphere. We have verified that, in
this way, the magnetosphere in our simulations obeys $\sigma :=B_i B^i/(4
\pi n m_e) \gg \Gamma$ ~\citep{Beskin1997}, with typical values of
$\sigma \gtrsim 100$ and $\sigma/\Gamma \gtrsim 10$.

We carried out an extensive campaign of GRPIC simulations, varying the
dimensionless spin parameter $a_*$ while keeping all other parameters
fixed (details and results are reported in~\citet{Meringolo2025a}). In
Fig.~~\ref{split} (left panel), we report a representative view of the
black hole magnetosphere for the most extreme configuration considered, 
namely a Kerr black hole with spin $a_* = 0.999$. The left panel shows
the magnetisation $\sigma$ at $\bar{t}/M = 14$, when the simulation has
reached a stationary state and a chain of plasmoids has fully formed. A
current sheet is clearly visible along the equatorial plane, where
magnetic reconnection occurs and $\sigma \lesssim 1$ due to the high
plasma density along the layer. Plasmoid formation is ubiquitous along
the current sheet: two large magnetic structures are visible at $r \sim
5\, M$, moving away from the black hole and merging to form a larger
plasmoid.

In the right panel, we instead show the alignment between the current
density and the magnetic field, $\gamma_{ij} J^i B^j / B_0^2$. As
expected, a strong alignment is present inside the closed field lines of
the plasmoids, with a polarity reversal across the equatorial plane. We
overplot magnetic-field lines (teal), highlighting that the
split-monopole configuration is strongly perturbed by plasmoid formation
and dynamics. Also shown in Fig.~\ref{split} is the region inside the
outer event horizon $r_+$ (black-shaded area) and the position of the
ergosphere, or static limit, (light-blue line).  
Within the inner light surface (shown with red line), massive particles 
cannot corotate and acquire instead large
radial infalling velocities at all latitudes, reaching Lorentz factors up
to $\Gamma \sim 200$ and comparable to those observed for the particles
accelerated by the magnetic reconnection along the equatorial plane.

\begin{figure*}[t]
  \centering
  \includegraphics[width=0.48\textwidth]{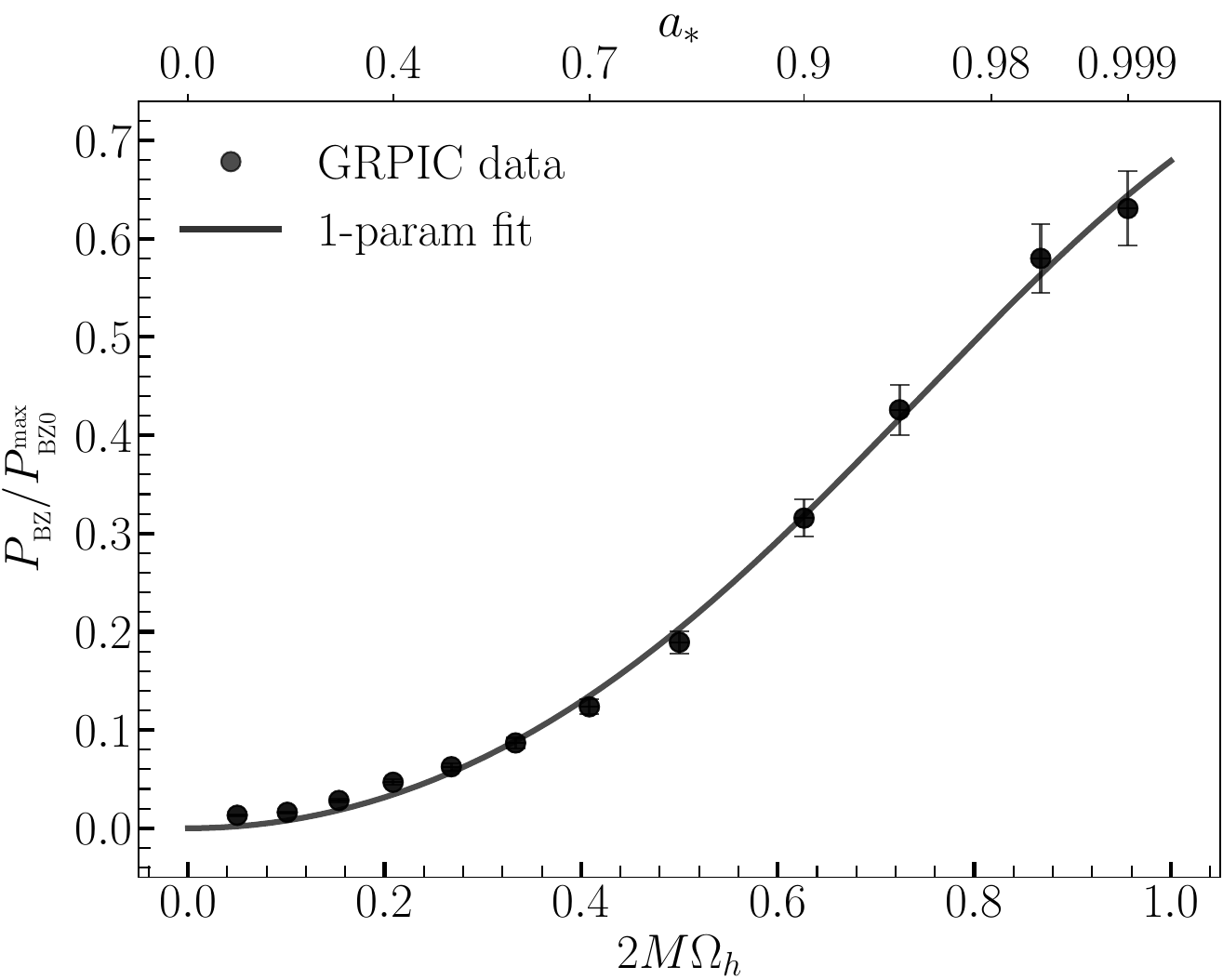}
  \hfill
  \includegraphics[width=0.48\textwidth]{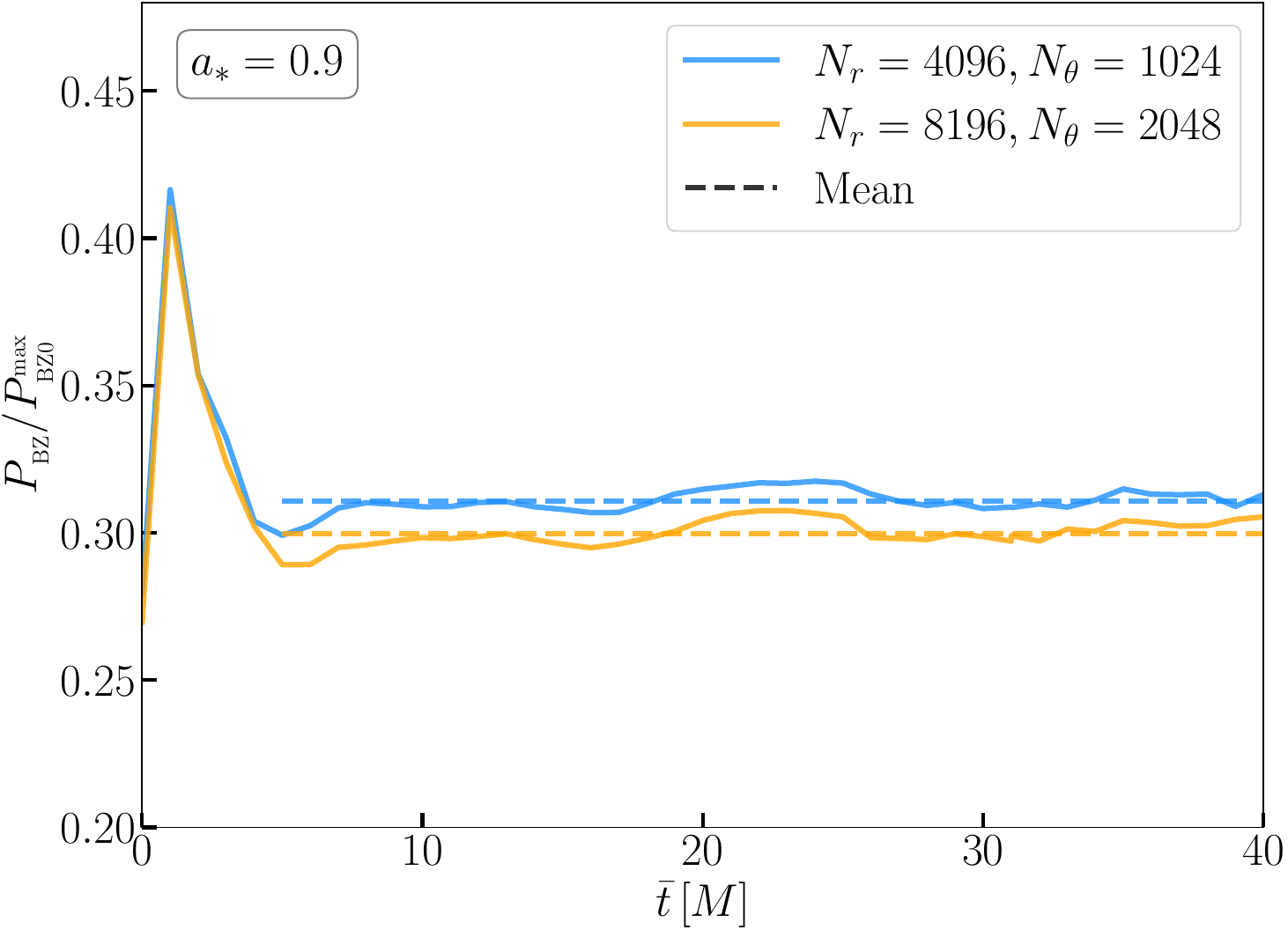}
  \caption{\textit{Left panel:} Normalized BZ luminosity, as a function
    of the black-hole angular velocity (see the top horizontal axis for a
    mapping in terms of the dimensionless spin of the black hole) for all
    of our GRPIC simulations (black filled circles) and with the
    associated numerical errors. The black solid line shows the analytic
    expression (Equation ~\eqref{eq:f_analytic}) after rescaling the
    factor $\kappa$ to account for the specific magnetic-field topology
    (one-parameter fit). \textit{Right panel:} Resolution test for the
    representative spin $a_*=0.9$, showing the time evolution of the
    normalized power. The number of points $N_r$ and $N_\theta$ are
    reported in the legend for each case, while dashed lines report the
    respective averages, computed in the time window $5 \leq \bar{t}/M
    \leq 40$.}
    \label{split2}
\end{figure*}

Using our large set of simulations, we have been able to measure the
power associated with the Blandford-Znajek (BZ) mechanism, which provides
an efficient conversion of black-hole rotational energy into
electromagnetic energy and operates in force-free black-hole
magnetospheres~\citep{Blandford1977}. We compute the power extracted via
the BZ mechanism by evaluating the Poynting flux through a two-sphere
located close to the event horizon, according to
\begin{equation}
  P_{_{\rm BZ}}=\int \left(T^r_{~~t}\right)_{_{\rm
      EM}}\sqrt{-g}\, d\theta\, d\phi = 2\pi \int_0^\pi
  S^r\sqrt{\gamma}\, d\theta\,, 
  \label{Pbz}
\end{equation}
where $\boldsymbol{T}_{_{\rm EM}}$ is the electromagnetic part of the
energy-momentum tensor, and $S^r := \left( E_\theta H_\phi - E_\phi
H_\theta \right) / (4 \pi \sqrt{\gamma})$ is the radial component of the
Poynting vector. The (time-averaged) results of our GRPIC simulations
are reported with black filled circles in the right panel of
Fig.~\ref{split2}, with the corresponding error-bars. Notice that we
normalise the powers to $P^{_{\rm max}}_{_{\rm BZ0}}\approx
10^{-3}(\Phi_h/M)^2$, representing the maximum value of the original BZ
estimate attained for $M\Omega_h = 1/2$, and where $\Phi_{h}$ is the
(half-hemisphere) magnetic flux~\citep{Meringolo2025a}. Expression
\eqref{Pbz} can be modelled generically as
\begin{equation}
  \label{eq:P_gen}
  P_{_{\rm BZ}}=\frac{\kappa}{4\pi}\, \Phi_{h}^2\, F(\Omega_{h})\,, 
\end{equation}
being $\kappa$ a scaling factor related to the topology of the
magnetosphere only, and $F(\Omega_h)$ a function of the spin of the black
hole. Recently, using analytic techniques that combine perturbation
theory with a matched-asymptotic expansion scheme~\citep{Armas2020, 
  Camilloni2022}, it was possible to obtain high-order corrections in
$F(\Omega_h)$ up to $\mathcal{O}(\Omega^8_h)$, namely
\begin{align}
\label{eq:f_analytic}
F_{\rm an}(\Omega_{h}) = \Omega^2_{h} \left[ 1 + \tilde{\alpha}
      (M\Omega_{h})^2 + \tilde{\beta} (M\Omega_{h})^4
      + \tilde{\gamma} |M\Omega_{h}|^5 \right. \nonumber \\
      + \left. \left(\tilde{\delta} + \tilde{\epsilon}
      \log|M\Omega_{h}|\right) (M\Omega_{h})^6\right] 
 + \mathcal{O}(M^9\Omega_{h}^9)\,.
\end{align}
The black solid line in Fig.~\ref{split2} reports the
expression~\eqref{eq:P_gen}, combined with the
dependence~\eqref{eq:f_analytic}, where a one-parameter fit was made to
fix the scaling factor $\kappa=0.041$.

In the right panel of Fig.~\ref{split2}, we show the time evolution of
$P_{\rm BZ}/P^{\rm max}_{\rm BZ0}$ for a representative simulation with
$a_* = 0.9$, while varying the grid resolution and keeping all other
parameters fixed. The light-blue line represents the standard resolution
we used in all our split-monopole campaign of simulations, whereas the
orange line shows the result of a simulation with twice the number of
grid points in both the $r$- and $\theta$-direction. We note that a peak
in the BZ power suddenly appears when the first particle injection starts
at $\bar{t}/M = 0$, perturbing the stationary electrovacuum
configuration. To avoid unphysical measurements, we compute the averaged
power only after stationarity is reached, by taking the mean value of
$P_{\rm BZ}/ P^{\rm max}_{\rm BZ0}$ in the time interval $5 \leq
\bar{t}/M \leq 40$. The dashed lines (with the corresponding colour code)
indicate the mean values for the two resolutions, which correspond to
normalized powers of $0.316$ and $0.299$, respectively, with a relative
difference of $\lesssim 6\%$.

\section{Conclusions}
\label{SecCon}

We have presented \texttt{FPIC}, a newly developed PIC code for the
solution of collisionless plasma in curved spacetimes and in the vicinity
of black holes. With the goal of ensuring the reproducibility of the
results obtained with \texttt{FPIC}, we have provided a detailed
discussion of the numerical methods employed to solve the equations of
motion of the particles and the evolution equations for the
electromagnetic fields.

While most of the methods employed are well-known and have been used in
other numerical codes of this type, we have also proposed a new hybrid
approach for the evolution of the equations of motion of the charged
particles. In the novel method, the time integrator is not kept fixed but
chosen in real time based on the evolution of the energy error, which is
monitored either via its time derivative or via its instantaneous
value. The key idea is simple yet effective: far from the black hole, the
robust and efficient RK4 integrator ensures a good energy conservation,
but loses precision as the particle nears the black-hole event
horizon. In such regions, a more accurate scheme -- such as the
Hamiltonian integrator -- would be preferable, but its computational
costs are considerably higher. An optimal compromise can be obtained by
switching between the two methods on the basis of the degree of curvature
of the spacetime. Furthermore, by taking advantage of the stability of
the Hamiltonian integrator at larger timesteps, an adaptive timestep is
employed when switching between the two schemes. The result of this
approach is a particle pusher that guarantees high precision at
comparatively small computational costs.

The numerical methods described in this way were then applied to a number
of testbeds aimed at validating the correctness of the code, testing its
robustness, and assessing its performance. The tests carried out
concentrated on the evolution of particles (neutral and charged) near
black holes, on the evolution of the electromagnetic fields in vacuum,
and on the combination of the two in genuine PIC simulations. More
specifically, within the first class of tests, we have considered the
dynamics of a neutral particle in the Schwarzschild and Kerr metrics, as
well as that of a charged particle embedded in a background Wald magnetic
field, in rotating black-hole spacetimes with different spins. Besides
reproducing well-known behaviours of the particles under the influence of
gravity and of a Lorentz force, the use of the hybrid approach provides
high precision with errors being dominated by the interpolation scheme
used to couple fields and particles. For the second class of test, on the
other hand, we have validated the code's ability to recover the vacuum
Wald solution for a rotating black hole when initialised with the
Schwarzschild Wald solution. In this case, a so-called Meissner effect
expels out the magnetic-field lines threading the horizon, reaching a
stationary regime.

Finally, we have tested the robustness of code with fully GRPIC scenarios
in which a pair plasma is evolved self-consistently with the
electromagnetic fields that are produced by its dynamics. In the case in
which the electrovacuum Wald solution is filled with a pair plasma, we
observed that the magnetic-field lines threading the ergosphere are
dragged into the black hole and form an equatorial current sheet within
the ergosphere. Electrons with negative energy at infinity are ubiquitous
inside the ergosphere along the current sheet, indicating that the
Penrose process is active there. In addition, we have presented results
for a rotating black hole threaded by a split-monopole magnetic
field. This configuration is well-known to generate a current sheet near
the equatorial plane and to trigger magnetic reconnection, which is
responsible for the production of chains of relativistic plasmoids
\citep[see][for details]{Meringolo2025a}. For these calculations we have
measured the power associated with the Blandford and Znajek mechanism,
founding excellent agreement with the high-order predictions coming from
perturbation theory. Because the match requires the calculation of a
single coefficient independent of the black-hole spin and related to the
specific magnetic-field topology, the excellent match is a very strong
validation test. 

In summary, we have presented in detail the numerical methods to be
implemented in the construction of a PIC code within a
general-relativistic framework and shown that a code built in this way,
\ie \texttt{FPIC} is to reproduce correctly and accurately fully
nonlinear plasma dynamics near rotating black holes. While a number of
applications await the use of this code, a number of extensions are
possible. First, we recall that we have adopted a simplified pair-plasma
injection scheme based on the local Goldreich-Julian density. We found
that this approach substantially speeds up the simulations while yielding
results that are both qualitatively and quantitatively consistent with
those obtained using more sophisticated and physically detailed injection
prescriptions \citep{Parfrey2019, ElMellah2021, Chen2025}. Nonetheless,
to achieve fully realistic magnetospheric modelling, more accurate
particle-injection strategies need to be developed and tested. Second, as
customary in PIC simulations, our approach ignores radiative processes,
which may however play a role under specific physical conditions and, in
particular, when the particle experience extremely large accelerations
near the black-hole horizon. Hence, future developments will have to
focus on implementing inverse-Compton emission, photon-photon
pair-production mechanisms, and radiation-reaction forces. Finally, but
not less importantly, to model ever more realistic astrophysical
scenarios, the code testing and validation needs to be extended to fully
3D configurations. Because the associated computational costs easily
become unsustainable when the calculations are performed with CPUs,
\texttt{FPIC} need to be extended with a hybrid MPI-OpenMP
parallelisation strategy that includes GPU acceleration through OpenMP
offloading. Progress in many of these directions is already on-going and
will be presented in future works.

\begin{acknowledgements}
It is a pleasure to acknowledge Bj\"orn Dick, Tobias Haas, Markus Kurz,
Hirav Patel, and Johanna Potyka, for their help and the fruitful
discussion at the hackathon \textit{Porting and Optimization for Hunter}
at HLRS, Stuttgart, Germany. We are also grateful to F. Camilloni for
useful discussions and suggestions. This research was supported by the
ERC Advanced Grant JETSET: Launching, propagation and emission of
relativistic jets from binary mergers and across mass scales (Grant
No. 884631). L.R. acknowledges the Walter Greiner Gesellschaft zur
F\"orderung der physikalischen Grundlagenforschung e.V. through the Carl
W. Fueck Laureatus Chair. The simulations were performed on HPE Apollo
HAWK at the High Performance Computing Center Stuttgart (HLRS) under the
grant BNSMIC, on SuperMUC-NG at the Leibniz Supercomputing Centre (LRZ)
under the grant MICMAC, and on the local ITP Supercomputing Clusters
Goethe-HLR, Iboga and Calea.
\end{acknowledgements}


\begin{appendix}
\onecolumn

\section{The Kerr metric}
\label{AppKS}

The spacetime around a stationary rotating black hole is described by the
Kerr metric~\citep{Kerr63}. In Kerr-Schild coordinates, the metric is
given by~\citep{Komissarov2004b}
\begin{equation}
\gamma_{ij}  = 
\begin{pmatrix}
 \xi  & 0 & -a_*^2 \, \xi\sin^2 \theta \\
 0   & \rho^2 & 0  \\
 -a_*^2 \, \xi \sin^2 \theta   & 0 &  \sin^2 \theta 
 \left[\rho^2 + a_*^2 \, \xi \sin^2 \theta \right] \,, \\
\end{pmatrix} 
\label{gsf}
\end{equation}
\begin{equation}
\gamma^{ij}  = 
\begin{pmatrix}
 \dfrac{a_*^2 \sin^2 \theta}{\rho^2} +\xi^{-1} & 0 & \dfrac{a_*}{\rho^2}
 \\ 0 & \dfrac{1}{\rho^2} & 0 \\ \dfrac{a_*}{\rho^2} & 0 &
 \dfrac{1}{\rho^2 \sin^2 \theta} \,, \\
\end{pmatrix} 
\label{gsf_2}
\end{equation}
where $\xi = 1+2Mr/\rho^2$ and $\rho^2 = r^2 + a_*^2 \cos^2 \theta$. The 
lapse function $\alpha$ and the shift vector $\beta^i = \beta^r$ read
\begin{equation} 
\alpha = \xi^{-1/2}\,, ~~~~~~~~~~~ \beta^r = \frac{2Mr}{\rho^2\, \xi } =
\frac{\xi-1}{\xi }\,, 
\end{equation}
meaning that the grid is moving towards the black hole. This motion of
the grid removes the well-known coordinate singularity at the event
horizon present in the more widely used Boyer-Lindquist coordinates, but
comes at the cost of a non-diagonal form of the spatial metric. The
metric determinant is represented by
\begin{equation}
\gamma = \rho^4 \, \xi \, \sin^2 \theta .
\label{det}
\end{equation} 
Note that the limit $a_* \rightarrow 0$ the above Kerr metric becomes
diagonal and it reduces to the Schwarzschild metric.

\section{Analytical initial conditions}

\subsection{The Wald solution}
\label{ICwald}

The stationary Wald field solution in the presence of a spinning black
hole and consistent with a uniform magnetic field of strength
$\tilde{B_0}$ in Kerr-Schild coordinates is given by
\begin{eqnarray}
B^r &=& \dfrac{\tilde{B_0}}{\sqrt{\gamma}} \left[ r^2 + a_*^2 - 2r +
  \dfrac{2r(r^4-a_*^4)}{\rho^4} \right] \sin \theta \cos \theta \,, 
\label{wald1}
\\
B^\theta &=& -\dfrac{\tilde{B_0}}{\sqrt{\gamma}} \left[ r +
  a_*^2(1+\cos^2 \theta) \dfrac{2r^2-\rho^2}{\rho^4} \right] \sin^2
\theta \,, \\
B^\varphi &=& -\dfrac{\tilde{B_0}}{\sqrt{\gamma}} \left[ \dfrac{2a_*r
    (a_*^2-r^2) }{\rho^4} -a_* \right] \sin \theta \cos \theta \,, \\
E_r &=& \tilde{B_0} \, a_* (1+\cos^2 \theta) \dfrac{a_*^2 \cos^2 \theta
  -r^2}{\rho^4} \,, \\
E_\theta &=& \tilde{B_0} \, a_* \dfrac{2 a_* r(a_*^2-r^2)}{\rho^4} \sin
\theta \cos \theta\,, \\ E_\varphi &=& 0\,, 
\label{wald6}
\end{eqnarray}
where $\tilde{B_0}=10^3$ and the four-potential is reported in
Eq.~(\ref{Amu}). From the covariant components of the electric field
$E_i$, one can obtain the FIDO's electric field $D_i$ by means of the
relation (\ref{Ei}), obtaining the contravariant components as $D^j =
\gamma^{ij} D_i$. We have adopted the above set of equations
(\ref{wald1})--(\ref{wald6}) as initial conditions for the $B^i$ and
$D^i$ fields in different cases: 1) in Section~\ref{testpart}, when
evolving the charged particles embedded in a magnetic field; 2) in
Section~\ref{waldvac}, for the vacuum Wald test by setting $a_*=0$ for
the Schwarzschild solution; 3) in Section~\ref{waldfull}, for the
plasma-filled Wald case.

\subsection{The split-monopole solution}
\label{ICsplit}

For the split monopole configuration presented in Section~\ref{SM}, we start 
from an electrovacuum solution in the Schwarzschild spacetime, initialised by 
means of a magnetic-flux function $\Psi(r,\theta) = 4M^2B_0 \, \zeta(\theta)
(1-\cos\theta)$. This standard split-construction is realised through a 
step-function $\zeta$ that introduces a current-sheet in the equatorial plane 
necessary to sustain the reversed polarity of the magnetic field \citep{Gralla2014}. 
In our simulations we considered an approximate expression 
$\zeta = 2/\pi \arctan[\zeta_0(\theta-\pi/2)]$.
After the injection, once the magnetosphere settles in plasma-filled conditions, 
the simulation approaches solutions obtained from perturbative studies of FFE in 
Kerr spacetime~\citep{Blandford1977,Tanabe2008, Armas2020, Camilloni2024}, namely

\begin{eqnarray}
\label{sp1}
B^r &=& -2 \tilde{B_0} \arctan\!\left[\dfrac{1}{2} \zeta_0 (\pi -2\theta)
  \right] \dfrac{1}{\pi\sqrt{r^3(2+r)}} \,, \\
B^\theta &=& 0 \,, \\
B^\varphi &=& -\tilde{B_0} \arctan\!\left[ \zeta_0
  (-\dfrac{\pi}{2}+\theta) \right] \dfrac{a_*(4+r)}{4 \pi\sqrt{r^5(2+r)}}
\,, \\
D^r &=& \tilde{B_0} \arctan\!\left[\dfrac{1}{2} \zeta_0 (\pi -2\theta)
  \right] \dfrac{8+r(4+r) }{4 \pi\sqrt{r^7(2+r)}} a_* \sin \theta \,, \\
D^\theta &=& 0\,, \\
D^\varphi &=& 0\,, \\
J^r &=& \tilde{B_0} \left\{ \arctan\!\left[\dfrac{1}{2} \zeta_0 (\pi -2\theta)
  \right] \cos \theta - \frac{2 \zeta_0 \sin \theta}{4+\zeta_0^2(\pi-2 \theta)^2} \right\}
  \dfrac{a_*}{2 \pi r^2 } \,, \\
J^\theta &=& 0 \, ,\\
J^\varphi &=& -\tilde{B_0} \dfrac{8 \zeta_0}{\pi \Big(4+\zeta_0^2 (\pi-2 \theta)^2\Big)  r^4 \sin \theta }  \,.
\label{sp2}
\end{eqnarray}
In our split monopole configuration we used Eqs.\,(\ref{sp1})--(\ref{sp2}) by setting 
$\tilde{B_0}=10^3$, $\zeta_0=10^3$ and $a_*=0$.

\section{Particle trajectories parameters}

Here we report physical and numerical parameters for the test particle
trajectories presented in Section~\ref{testpart}. We split the data in
Tab.~\ref{tab:params} in neutral-particle trajectories (top) and
charged-particle trajectories (bottom). For the charged particles we used
the same naming conventions and values reported in the literature
\citep{Levinson2018, Bacchini2019, Chen2025}. We also report in the
Tab.~\ref{tab:params_hyb} the numerical parameters used in our hybrid
approach for neutral particles, presented in the
Subsection~\ref{sec_hyb}.
 
\begin{table*}[h]
    \centering
    \caption{Initial conditions and parameters for particle trajectories
      reported in Subsections~\ref{sec_neutralpart} (neutral particles, 
      top rows) and~\ref{sec_chargedpart} (charged particles, bottom
      rows). Starting from the left, the columns represent: The name of
      simulation, the dimensionless spin $a_*$, the initial
      magnetic-field parameter $B_0$, the orbit's energy $E$ and angular
      momentum $L$, the initial particle position $\boldsymbol{r}$ and
      four-velocity $\boldsymbol{u}$. In all the cases we initialised
      $\varphi^0=0$.}
    \label{tab:params}
    \begin{tabular}{l c c c c c c}
        \hline \hline
        Name & $a_*$ & $B_0$ & $E$ & $L$ & $\{r^0, \theta^0\}$ &$\{u_{r}^0, u_{\theta}^0, u_\varphi^0\}$ \\
        \hline
        \multicolumn{7}{c}{Neutral particles} \\  
        \hline
        $\texttt{a.}0$ & 0.0   & 0.0 & 0.987649 & 3.9   & $\{73.239804, \pi/2\}$ & $\{0.027724, 0.0, 3.9\}$\\
        $\texttt{a.995}\_1$   & 0.995 & 0.0 & 0.916235 & 2.0   & $\{10.021533, \pi/2\}$ & $\{0.201211, 0.0, 2.0\}$\\
        $\texttt{a.995}\_2$   & 0.995 & 0.0 & 0.841423 & 1.82  & $\{4.376281, \pi/2\}$  & $\{0.487625, 0.0, 1.82\}$\\
        \hline
        \multicolumn{7}{c}{Charged particles} \\  
        \hline
        RKA2      & 0.7   & 2.0 & 1.81     & 16.25 & $\{4.2, \pi/2-0.1\}$   & $\{0.921532, 0.0, -1.93\}$\\
        RKA3      & 0.9   & 2.0 & 2.2      & 18.1  & $\{4.0, \pi/2-0.2\}$   & $\{0.638604, 0.0, 1.565\}$\\
        RKA8      & 0.9   & 1.0 & 1.24     & 5.0   & $\{3.0, \pi/2\}$       & $\{1.15912, 0.497, 0.365\}$\\
        \hline \hline
    \end{tabular}
\end{table*}

\begin{table}[h]
    \centering
    \caption{Numerical parameters for the hybrid integrator shown in
      Subsection~\ref{sec_hyb}. In the top rows of the table we report
      the simulation \texttt{a}.0, while in the bottom rows the two
      simulations \texttt{a.995}\_1 and \texttt{a.995}\_2. Starting from
      the left, we report: The integrator scheme, the associated $\Delta
      t_{\#}$ for the time integration, the value of the constraint
      $(\partial_t \widetilde{H})_{\text{max}}$ for the ``Hyb1'' case,
      and the value of the constraint $\widetilde{H}_\text{max}$ for the
      ``Hyb2'' case. }
    \label{tab:params_hyb}
    \begin{tabular}{l c c c c}
        \hline \hline
        Scheme & $\Delta t_{\text{RK4}}$ & $\Delta t_{\text{Ham}}$ & $(\partial_t \widetilde{H})_{\text{max}}$ & $\widetilde{H}_\text{max}$ \\
        \hline
        \multicolumn{5}{c}{\texttt{a.}0} \\  
        \hline
        RK4 & 0.05 & - & - & -    \\
        Ham & - & 1.0 & - & -    \\
        Hyb1 & 0.08 & 1.0 & 1.5 & -    \\
        Hyb2 & 0.5 & 1.0 & - & $10^{-15}$  \\
        \hline
        \multicolumn{5}{c}{\texttt{a.995\_1}, \texttt{a.995\_2}} \\  
        \hline
        RK4 & 0.05 & - & - & -    \\
        Ham & - & 0.1 & - & -    \\
        Hyb1 & 0.05 & 0.1 & 0.5 & -    \\
        Hyb2 & 0.05 & 0.1 & - & $10^{-12}$ \\
        \hline \hline
    \end{tabular}
\end{table}

\section{Discrete Hamiltonian integrator}
\label{ham_eq}

Starting from Equation~\eqref{Ham} and considering a particle in
electrovacuum, \ie without the influence of the Lorenz force, the
Hamiltonian reduces to
\begin{equation}
H(x^i, u_j) = \alpha \sqrt{1+\gamma^{lm} u_l u_m} - \beta^r u_r.
\end{equation}
From the above expression, considering axisymmetry and Kerr-Schild
coordinates, the discretized Hamiltonian equations can be written as
\citep{Bacchini2018}:

\begin{figure*}[h]
\begin{equation}
\begin{split}
\frac{r^{n+1}-r^n}{\Delta t} = \frac{1}{6} \Bigg\{
& \alpha(r^{n+1}, \theta^n)
\frac{
\gamma^{rr}(r^{n+1}, \theta^n)\big[u_r^{n+1}+u_r^{n}\big]
+ 2 \gamma^{r\varphi}(r^{n+1}, \theta^{n}) u_\varphi^{n}
}{
\sqrt{1+\xi_r^{n+1}(u_r^{n+1}, u_\theta^{n}, u_\varphi^{n})}
+ \sqrt{1+\xi_r^{n}(u_r^{n}, u_\theta^{n}, u_\varphi^{n})}
}
- \beta^r(r^{n+1}, \theta^n)
\\ 
+& \alpha(r^{n+1}, \theta^{n+1})
\frac{
\gamma^{rr}(r^{n+1}, \theta^{n+1})\big[u_r^{n+1}+u_r^{n}\big]
+ 2 \gamma^{r\varphi}(r^{n+1}, \theta^{n+1}, \varphi^{n+1}) u_\varphi^{n+1}
}{
\sqrt{1+\xi_r^{n+1}(u_r^{n+1}, u_\theta^{n+1}, u_\varphi^{n+1})}
+ \sqrt{1+\xi_r^{n}(u_r^{n}, u_\theta^{n+1}, u_\varphi^{n+1})}
}
- \beta^r(r^{n+1}, \theta^{n+1})
\\
+& \alpha(r^{n}, \theta^{n+1})
\frac{
\gamma^{rr}(r^{n}, \theta^{n+1})\big[u_r^{n+1}+u_r^{n}\big]
+ 2 \gamma^{r\varphi}(r^{n}, \theta^{n+1}, \varphi^{n+1}) u_\varphi^{n+1}
}{
\sqrt{1+\xi_r^{n+1}(u_r^{n+1}, u_\theta^{n+1}, u_\varphi^{n+1})}
+ \sqrt{1+\xi_r^{n}(u_r^{n}, u_\theta^{n+1}, u_\varphi^{n+1})}
}
- \beta^r(r^{n}, \theta^{n+1})
\\
+& \alpha(r^{n+1}, \theta^{n})
\frac{
\gamma^{rr}(r^{n+1}, \theta^{n})\big[u_r^{n+1}+u_r^{n}\big]
+ 2 \gamma^{r\varphi}(r^{n+1}, \theta^{n}, \varphi^{n+1}) u_\varphi^{n+1}
}{
\sqrt{1+\xi_r^{n+1}(u_r^{n+1}, u_\theta^{n}, u_\varphi^{n+1})}
+ \sqrt{1+\xi_r^{n}(u_r^{n}, u_\theta^{n}, u_\varphi^{n+1})}
}
- \beta^r(r^{n+1}, \theta^{n})
\\
+& \alpha(r^{n}, \theta^n)
\frac{
\gamma^{rr}(r^{n}, \theta^n)\big[u_r^{n}+u_r^{n}\big]
+ 2 \gamma^{r\varphi}(r^{n}, \theta^{n}, \varphi^n) u_\varphi^{n}
}{
\sqrt{1+\xi_r^{n+1}(u_r^{n+1}, u_\theta^{n}, u_\varphi^{n})}
+ \sqrt{1+\xi_r^{n}(u_r^{n}, u_\theta^{n}, u_\varphi^{n})}
}
- \beta^r(r^{n}, \theta^n)
\\
+& \alpha(r^{n}, \theta^n)
\frac{
\gamma^{rr}(r^{n}, \theta^n)\big[u_r^{n}+u_r^{n}\big]
+ 2 \gamma^{r\varphi}(r^{n}, \theta^{n}, \varphi^{n+1}) u_\varphi^{n+1}
}{
\sqrt{1+\xi_r^{n}(u_r^{n+1}, u_\theta^{n+1}, u_\varphi^{n+1})}
+ \sqrt{1+\xi_r^{n}(u_r^{n}, u_\theta^{n+1}, u_\varphi^{n+1})}
}
- \beta^r(r^{n}, \theta^n)
\Bigg\},
\end{split}
\end{equation}
\end{figure*}

\begin{figure*}[h]
\begin{equation}
\begin{split}
\frac{\theta^{n+1}-\theta^n}{\Delta t} = \frac{1}{6} \Bigg\{
& \alpha(r^{n+1}, \theta^{n+1})
\frac{
\gamma^{\theta \theta}(r^{n+1}, \theta^{n+1})\big[u_\theta^{n+1}+u_\theta^{n}\big]
}{
\sqrt{1+\xi_\theta^{n+1}(u_r^{n+1}, u_\theta^{n+1}, u_\varphi^{n})}
+ \sqrt{1+\xi_\theta^{n}(u_r^{n+1}, u_\theta^{n}, u_\varphi^{n})}
}
\\ 
+& \alpha(r^{n}, \theta^{n+1})
\frac{
\gamma^{\theta \theta}(r^{n}, \theta^{n+1})\big[u_\theta^{n+1}+u_\theta^{n}\big]
}{
\sqrt{1+\xi_\theta^{n+1}(u_r^{n}, u_\theta^{n+1}, u_\varphi^{n})}
+ \sqrt{1+\xi_\theta^{n}(u_r^{n}, u_\theta^{n}, u_\varphi^{n})}
}
\\ 
+& \alpha(r^{n}, \theta^{n})
\frac{
\gamma^{\theta \theta}(r^{n}, \theta^{n})\big[u_\theta^{n+1}+u_\theta^{n}\big]
}{
\sqrt{1+\xi_\theta^{n+1}(u_r^{n}, u_\theta^{n+1}, u_\varphi^{n})}
+ \sqrt{1+\xi_\theta^{n}(u_r^{n}, u_\theta^{n}, u_\varphi^{n})}
}
\\ 
+& \alpha(r^{n+1}, \theta^{n+1})
\frac{
\gamma^{\theta \theta}(r^{n+1}, \theta^{n+1})\big[u_\theta^{n+1}+u_\theta^{n}\big]
}{
\sqrt{1+\xi_\theta^{n+1}(u_r^{n+1}, u_\theta^{n+1}, u_\varphi^{n+1})}
+ \sqrt{1+\xi_\theta^{n}(u_r^{n+1}, u_\theta^{n}, u_\varphi^{n+1})}
}
\\ 
+& \alpha(r^{n+1}, \theta^{n})
\frac{
\gamma^{\theta \theta}(r^{n+1}, \theta^{n})\big[u_\theta^{n+1}+u_\theta^{n}\big]
}{
\sqrt{1+\xi_\theta^{n+1}(u_r^{n+1}, u_\theta^{n+1}, u_\varphi^{n})}
+ \sqrt{1+\xi_\theta^{n}(u_r^{n+1}, u_\theta^{n}, u_\varphi^{n})}
}
\\ 
+& \alpha(r^{n+1}, \theta^{n})
\frac{
\gamma^{\theta \theta}(r^{n+1}, \theta^{n})\big[u_\theta^{n+1}+u_\theta^{n}\big]
}{
\sqrt{1+\xi_\theta^{n+1}(u_r^{n+1}, u_\theta^{n+1}, u_\varphi^{n+1})}
+ \sqrt{1+\xi_\theta^{n}(u_r^{n+1}, u_\theta^{n}, u_\varphi^{n+1})}
}
\Bigg\},
\end{split}
\end{equation}
\end{figure*}

\begin{figure*}[h]
\begin{equation}
\begin{split}
\frac{\varphi^{n+1}-\varphi^n}{\Delta t} = \frac{1}{6} \Bigg\{
& \alpha(r^{n+1}, \theta^{n+1})
\frac{
2 \gamma^{r \varphi}(r^{n+1}, \theta^{n+1}) u_r^{n+1} +  \gamma^{\varphi \varphi}(r^{n+1}, \theta^{n+1}) \big[u_\varphi^{n+1}+u_\varphi^{n}\big]
}{
\sqrt{1+\xi_\varphi^{n+1}(u_r^{n+1}, u_\theta^{n+1}, u_\varphi^{n+1})}
+ \sqrt{1+\xi_\varphi^{n}(u_r^{n+1}, u_\theta^{n+1}, u_\varphi^{n})}
}
\\ 
+& \alpha(r^{n}, \theta^{n+1})
\frac{
2 \gamma^{r \varphi}(r^{n}, \theta^{n+1}) u_r^{n} +  \gamma^{\varphi \varphi}(r^{n}, \theta^{n+1}) \big[u_\varphi^{n+1}+u_\varphi^{n}\big]
}{
\sqrt{1+\xi_\varphi^{n+1}(u_r^{n}, u_\theta^{n+1}, u_\varphi^{n+1})}
+ \sqrt{1+\xi_\varphi^{n}(u_r^{n}, u_\theta^{n+1}, u_\varphi^{n})}
}
\\ 
+& \alpha(r^{n}, \theta^{n+1})
\frac{
2 \gamma^{r \varphi}(r^{n}, \theta^{n+1}) u_r^{n} +  \gamma^{\varphi \varphi}(r^{n}, \theta^{n+1}) \big[u_\varphi^{n+1}+u_\varphi^{n}\big]
}{
\sqrt{1+\xi_\varphi^{n+1}(u_r^{n}, u_\theta^{n+1}, u_\varphi^{n+1})}
+ \sqrt{1+\xi_\varphi^{n}(u_r^{n}, u_\theta^{n+1}, u_\varphi^{n})}
}
\\ 
+& \alpha(r^{n}, \theta^{n})
\frac{
2 \gamma^{r \varphi}(r^{n}, \theta^{n}) u_r^{n} +  \gamma^{\varphi \varphi}(r^{n}, \theta^{n}) \big[u_\varphi^{n+1}+u_\varphi^{n}\big]
}{
\sqrt{1+\xi_\varphi^{n+1}(u_r^{n}, u_\theta^{n}, u_\varphi^{n+1})}
+ \sqrt{1+\xi_\varphi^{n}(u_r^{n}, u_\theta^{n}, u_\varphi^{n})}
}
\\ 
+& \alpha(r^{n+1}, \theta^{n+1})
\frac{
2 \gamma^{r \varphi}(r^{n+1}, \theta^{n+1}) u_r^{n+1} +  \gamma^{\varphi \varphi}(r^{n+1}, \theta^{n+1}) \big[u_\varphi^{n+1}+u_\varphi^{n}\big]
}{
\sqrt{1+\xi_\varphi^{n+1}(u_r^{n+1}, u_\theta^{n+1}, u_\varphi^{n+1})}
+ \sqrt{1+\xi_\varphi^{n}(u_r^{n+1}, u_\theta^{n+1}, u_\varphi^{n})}
}
\\ 
& \alpha(r^{n}, \theta^{n})
\frac{
2 \gamma^{r \varphi}(r^{n}, \theta^{n}) u_r^{n} +  \gamma^{\varphi \varphi}(r^{n}, \theta^{n}) \big[u_\varphi^{n+1}+u_\varphi^{n}\big]
}{
\sqrt{1+\xi_\varphi^{n+1}(u_r^{n}, u_\theta^{n}, u_\varphi^{n+1})}
+ \sqrt{1+\xi_\varphi^{n}(u_r^{n}, u_\theta^{n}, u_\varphi^{n})}
}
\Bigg\},
\end{split}
\end{equation}
\end{figure*}

\begin{figure*}[h]
\begin{equation}
\begin{split}
\label{ur1}
\frac{u_r^{n+1}-u_r^n}{\Delta t} &= \frac{1}{6} \Bigg\{
- \dfrac{1}{2} \Bigg(
      \sqrt{1+\zeta_r^{n+1}(u_r^{n}, u_\theta^{n}, u_\varphi^{n})} 
    + \sqrt{1+\zeta_r^{n}(u_r^{n}, u_\theta^{n}, u_\varphi^{n})} 
  \Bigg) \, \partial_r \alpha(r^{n+1}, \theta^{n}) + u^n_r \partial_{r} \beta^r(r^{n+1}, \theta^{n})
\\
&-  \dfrac{
        \Big[ \alpha(r^{n+1}, \theta^{n}) + \alpha(r^{n}, \theta^{n}) \Big] 
        \Big[(u^n_r)^2 \partial_r \gamma^{rr} (r^{n+1}, \theta^{n})
           + (u^n_\theta)^2 \partial_r \gamma^{\theta \theta}
           + (u^n_\varphi)^2 \partial_r \gamma^{\varphi \varphi}
           + 2u^n_r u^n_\varphi \partial_r \gamma^{r \varphi}\Big]
      }{
        2\sqrt{1+\zeta_r^{n+1}(u_r^{n}, u_\theta^{n}, u_\varphi^{n})} + \sqrt{1+\zeta_r^{n}(u_r^{n}, u_\theta^{n}, u_\varphi^{n})}
      } 
\\ 
&- \dfrac{1}{2} \Bigg(
      \sqrt{1+\zeta_r^{n+1}(u_r^{n}, u_\theta^{n+1}, u_\varphi^{n+1})} 
    + \sqrt{1+\zeta_r^{n}(u_r^{n}, u_\theta^{n+1}, u_\varphi^{n+1})} 
  \Bigg) \, \partial_r \alpha(r^{n+1}, \theta^{n+1}) + u^{n}_r \partial_{r} \beta^r(r^{n+1}, \theta^{n+1})
\\
&-  \dfrac{
        \Big[ \alpha(r^{n+1}, \theta^{n+1}) + \alpha(r^{n}, \theta^{n+1}) \Big] 
        \Big[(u^n_r)^2 \partial_r \gamma^{rr} (r^{n+1}, \theta^{n+1})
           + (u^{n+1}_\theta)^2 \partial_r \gamma^{\theta \theta}
           + (u^{n+1}_\varphi)^2 \partial_r \gamma^{\varphi \varphi}
           + 2u^n_r u^{n+1}_\varphi \partial_r \gamma^{r \varphi}\Big]
      }{
        2\sqrt{1+\zeta_r^{n+1}(u_r^{n}, u_\theta^{n+1}, u_\varphi^{n+1})} + \sqrt{1+\zeta_r^{n}(u_r^{n}, u_\theta^{n+1}, u_\varphi^{n+1})}
      }
\\ 
&- \dfrac{1}{2} \Bigg(
      \sqrt{1+\zeta_r^{n+1}(u_r^{n+1}, u_\theta^{n+1}, u_\varphi^{n+1})} 
    + \sqrt{1+\zeta_r^{n}(u_r^{n+1}, u_\theta^{n+1}, u_\varphi^{n+1})} 
  \Bigg) \, \partial_r \alpha(r^{n+1}, \theta^{n+1}) + u^{n}_r \partial_{r} \beta^r(r^{n+1}, \theta^{n+1})
\\
&-  \dfrac{
        \Big[ \alpha(r^{n+1}, \theta^{n+1}) + \alpha(r^{n}, \theta^{n+1}) \Big] 
        \Big[(u^{n+1}_r)^2 \partial_r \gamma^{rr} (r^{n+1}, \theta^{n+1})
           + (u^{n+1}_\theta)^2 \partial_r \gamma^{\theta \theta}
           + (u^{n+1}_\varphi)^2 \partial_r \gamma^{\varphi \varphi}
           + 2u^{n+1}_r u^{n+1}_\varphi \partial_r \gamma^{r \varphi}\Big]
      }{
        2\sqrt{1+\zeta_r^{n+1}(u_r^{n+1}, u_\theta^{n+1}, u_\varphi^{n+1})} + \sqrt{1+\zeta_r^{n+1}(u_r^{n}, u_\theta^{n+1}, u_\varphi^{n+1})}
      }
\\ 
&- \dfrac{1}{2} \Bigg(
      \sqrt{1+\zeta_r^{n+1}(u_r^{n}, u_\theta^{n}, u_\varphi^{n+1})} 
    + \sqrt{1+\zeta_r^{n}(u_r^{n}, u_\theta^{n}, u_\varphi^{n+1})} 
  \Bigg) \, \partial_r \alpha(r^{n+1}, \theta^{n}) + u^{n}_r \partial_{r} \beta^r(r^{n+1}, \theta^{n})
\\
&-  \dfrac{
        \Big[ \alpha(r^{n+1}, \theta^{n}) + \alpha(r^{n}, \theta^{n}) \Big] 
        \Big[(u^n_r)^2 \partial_r \gamma^{rr} (r^{n+1}, \theta^{n})
           + (u^{n}_\theta)^2 \partial_r \gamma^{\theta \theta}
           + (u^{n+1}_\varphi)^2 \partial_r \gamma^{\varphi \varphi}
           + 2u^n_r u^{n+1}_\varphi \partial_r \gamma^{r \varphi}\Big]
      }{
        2\sqrt{1+\zeta_r^{n+1}(u_r^{n}, u_\theta^{n}, u_\varphi^{n+1})} + \sqrt{1+\zeta_r^{n}(u_r^{n}, u_\theta^{n}, u_\varphi^{n+1})}
      }
\\ 
&- \dfrac{1}{2} \Bigg(
      \sqrt{1+\zeta_r^{n+1}(u_r^{n+1}, u_\theta^{n}, u_\varphi^{n})} 
    + \sqrt{1+\zeta_r^{n}(u_r^{n+1}, u_\theta^{n}, u_\varphi^{n})} 
  \Bigg) \, \partial_r \alpha(r^{n+1}, \theta^{n}) + u^{n}_r \partial_{r} \beta^r(r^{n+1}, \theta^{n})
\\
&-  \dfrac{
        \Big[ \alpha(r^{n+1}, \theta^{n}) + \alpha(r^{n}, \theta^{n}) \Big] 
        \Big[(u^{n+1}_r)^2 \partial_r \gamma^{rr} (r^{n+1}, \theta^{n})
           + (u^{n}_\theta)^2 \partial_r \gamma^{\theta \theta}
           + (u^{n}_\varphi)^2 \partial_r \gamma^{\varphi \varphi}
           + 2u^{n+1}_r u^{n}_\varphi \partial_r \gamma^{r \varphi}\Big]
      }{
        2\sqrt{1+\zeta_r^{n+1}(u_r^{n+1}, u_\theta^{n}, u_\varphi^{n})} + \sqrt{1+\zeta_r^{n+1}(u_r^{n+1}, u_\theta^{n}, u_\varphi^{n})}
      }
\\ 
&- \dfrac{1}{2} \Bigg(
      \sqrt{1+\zeta_r^{n+1}(u_r^{n+1}, u_\theta^{n}, u_\varphi^{n+1})} 
    + \sqrt{1+\zeta_r^{n}(u_r^{n+1}, u_\theta^{n}, u_\varphi^{n+1})} 
  \Bigg) \, \partial_r \alpha(r^{n+1}, \theta^{n}) + u^{n}_r \partial_{r} \beta^r(r^{n+1}, \theta^{n})
\\
&-  \dfrac{
        \Big[ \alpha(r^{n+1}, \theta^{n}) + \alpha(r^{n}, \theta^{n}) \Big] 
        \Big[(u^{n+1}_r)^2 \partial_r \gamma^{rr} (r^{n+1}, \theta^{n})
           + (u^{n}_\theta)^2 \partial_r \gamma^{\theta \theta}
           + (u^{n+1}_\varphi)^2 \partial_r \gamma^{\varphi \varphi}
           + 2u^{n+1}_r u^{n+1}_\varphi \partial_r \gamma^{r \varphi}\Big]
      }{
        2\sqrt{1+\zeta_r^{n+1}(u_r^{n+1}, u_\theta^{n}, u_\varphi^{n+1})} + \sqrt{1+\zeta_r^{n}(u_r^{n+1}, u_\theta^{n}, u_\varphi^{n+1})}
      }
\Bigg\},
\end{split}
\end{equation}
\end{figure*}

\begin{figure*}[h]
\begin{equation}
\begin{split}
\label{ut1}
\frac{u_\theta^{n+1}-u_\theta^n}{\Delta t} &= \frac{1}{6} \Bigg\{
- \dfrac{1}{2} \Bigg(
      \sqrt{1+\zeta_\theta^{n+1}(u_r^{n+1}, u_\theta^{n}, u_\varphi^{n})} 
    + \sqrt{1+\zeta_\theta^{n}(u_r^{n+1}, u_\theta^{n}, u_\varphi^{n})} 
  \Bigg) \, \partial_\theta \alpha(r^{n+1}, \theta^{n}) + u^{n+1}_r \partial_\theta \beta^r(r^{n+1}, \theta^{n})
\\
&-  \dfrac{
        \Big[ \alpha(r^{n+1}, \theta^{n+1}) + \alpha(r^{n+1}, \theta^{n}) \Big] 
        \Big[(u^{n+1}_r)^2 \partial_r \gamma^{rr} (r^{n+1}, \theta^{n})
           + (u^n_\theta)^2 \partial_\theta \gamma^{\theta \theta}
           + (u^n_\varphi)^2 \partial_\theta \gamma^{\varphi \varphi}
           + 2u^{n+1}_r u^n_\varphi \partial_\theta \gamma^{r \varphi}\Big]
      }{
        2\sqrt{1+\zeta_\theta^{n+1}(u_r^{n+1}, u_\theta^{n}, u_\varphi^{n})} + \sqrt{1+\zeta_\theta^{n}(u_r^{n+1}, u_\theta^{n}, u_\varphi^{n})}
      } 
\\ 
&- \dfrac{1}{2} \Bigg(
      \sqrt{1+\zeta_\theta^{n+1}(u_r^{n}, u_\theta^{n}, u_\varphi^{n})} 
    + \sqrt{1+\zeta_\theta^{n}(u_r^{n}, u_\theta^{n}, u_\varphi^{n})} 
  \Bigg) \, \partial_\theta \alpha(r^{n}, \theta^{n+1}) + u^{n}_r \partial_\theta \beta^r(r^{n}, \theta^{n+1})
\\
&-  \dfrac{
        \Big[ \alpha(r^{n+1}, \theta^{n+1}) + \alpha(r^{n}, \theta^{n}) \Big] 
        \Big[(u^{n}_r)^2 \partial_r \gamma^{rr} (r^{n}, \theta^{n+1})
           + (u^n_\theta)^2 \partial_\theta \gamma^{\theta \theta}
           + (u^n_\varphi)^2 \partial_\theta \gamma^{\varphi \varphi}
           + 2u^{n}_r u^n_\varphi \partial_\theta \gamma^{r \varphi}\Big]
      }{
        2\sqrt{1+\zeta_\theta^{n+1}(u_r^{n}, u_\theta^{n}, u_\varphi^{n})} + \sqrt{1+\zeta_\theta^{n}(u_r^{n}, u_\theta^{n}, u_\varphi^{n})}
      } 
\\ 
&- \dfrac{1}{2} \Bigg(
      \sqrt{1+\zeta_\theta^{n+1}(u_r^{n}, u_\theta^{n+1}, u_\varphi^{n})} 
    + \sqrt{1+\zeta_\theta^{n}(u_r^{n}, u_\theta^{n+1}, u_\varphi^{n})} 
  \Bigg) \, \partial_\theta \alpha(r^{n}, \theta^{n+1}) + u^{n}_r \partial_\theta \beta^r(r^{n}, \theta^{n+1})
\\
&-  \dfrac{
        \Big[ \alpha(r^{n+1}, \theta^{n+1}) + \alpha(r^{n}, \theta^{n}) \Big] 
        \Big[(u^{n}_r)^2 \partial_r \gamma^{rr} (r^{n}, \theta^{n+1})
           + (u^{n+1}_\theta)^2 \partial_\theta \gamma^{\theta \theta}
           + (u^n_\varphi)^2 \partial_\theta \gamma^{\varphi \varphi}
           + 2u^{n}_r u^n_\varphi \partial_\theta \gamma^{r \varphi}\Big]
      }{
        2\sqrt{1+\zeta_\theta^{n+1}(u_r^{n}, u_\theta^{n+1}, u_\varphi^{n})} + \sqrt{1+\zeta_\theta^{n}(u_r^{n}, u_\theta^{n+1}, u_\varphi^{n})}
      } 
\\ 
&- \dfrac{1}{2} \Bigg(
      \sqrt{1+\zeta_\theta^{n+1}(u_r^{n+1}, u_\theta^{n}, u_\varphi^{n+1})} 
    + \sqrt{1+\zeta_\theta^{n}(u_r^{n+1}, u_\theta^{n}, u_\varphi^{n+1})} 
  \Bigg) \, \partial_\theta \alpha(r^{n+1}, \theta^{n+1}) + u^{n+1}_r \partial_\theta \beta^r(r^{n+1}, \theta^{n+1})
\\
&-  \dfrac{
        \Big[ \alpha(r^{n+1}, \theta^{n+1}) + \alpha(r^{n+1}, \theta^{n}) \Big] 
        \Big[(u^{n+1}_r)^2 \partial_r \gamma^{rr} (r^{n+1}, \theta^{n+1})
           + (u^n_\theta)^2 \partial_\theta \gamma^{\theta \theta}
           + (u^{n+1}_\varphi)^2 \partial_\theta \gamma^{\varphi \varphi}
           + 2u^{n+1}_r u^{n+1}_\varphi \partial_\theta \gamma^{r \varphi}\Big]
      }{
        2\sqrt{1+\zeta_\theta^{n+1}(u_r^{n+1}, u_\theta^{n}, u_\varphi^{n+1})} + \sqrt{1+\zeta_\theta^{n}(u_r^{n+1}, u_\theta^{n}, u_\varphi^{n+1})}
      } 
\\ 
&- \dfrac{1}{2} \Bigg(
      \sqrt{1+\zeta_\theta^{n+1}(u_r^{n+1}, u_\theta^{n+1}, u_\varphi^{n})} 
    + \sqrt{1+\zeta_\theta^{n}(u_r^{n+1}, u_\theta^{n+1}, u_\varphi^{n})} 
  \Bigg) \, \partial_\theta \alpha(r^{n+1}, \theta^{n+1}) + u^{n+1}_r \partial_\theta \beta^r(r^{n+1}, \theta^{n+1})
\\
&-  \dfrac{
        \Big[ \alpha(r^{n+1}, \theta^{n+1}) + \alpha(r^{n+1}, \theta^{n}) \Big] 
        \Big[(u^{n+1}_r)^2 \partial_r \gamma^{rr} (r^{n+1}, \theta^{n+1})
           + (u^{n+1}_\theta)^2 \partial_\theta \gamma^{\theta \theta}
           + (u^{n+1}_\varphi)^2 \partial_\theta \gamma^{\varphi \varphi}
           + 2u^{n+1}_r u^{n}_\varphi \partial_\theta \gamma^{r \varphi}\Big]
      }{
        2\sqrt{1+\zeta_\theta^{n+1}(u_r^{n+1}, u_\theta^{n+1}, u_\varphi^{n})} + \sqrt{1+\zeta_\theta^{n}(u_r^{n+1}, u_\theta^{n+1}, u_\varphi^{n})}
      } 
\\ 
&- \dfrac{1}{2} \Bigg(
      \sqrt{1+\zeta_\theta^{n+1}(u_r^{n+1}, u_\theta^{n+1}, u_\varphi^{n+1})} 
    + \sqrt{1+\zeta_\theta^{n+1}(u_r^{n+1}, u_\theta^{n+1}, u_\varphi^{n+1})} 
  \Bigg) \, \partial_\theta \alpha(r^{n+1}, \theta^{n+1}) + u^{n+1}_r \partial_\theta \beta^r(r^{n+1}, \theta^{n+1})
\\
&-  \dfrac{
        \Big[ \alpha(r^{n+1}, \theta^{n+1}) + \alpha(r^{n+1}, \theta^{n}) \Big] 
        \Big[(u^{n+1}_r)^2 \partial_r \gamma^{rr} (r^{n+1}, \theta^{n+1})
           + (u^{n+1}_\theta)^2 \partial_\theta \gamma^{\theta \theta}
           + (u^{n+1}_\varphi)^2 \partial_\theta \gamma^{\varphi \varphi}
           + 2u^{n+1}_r u^{n+1}_\varphi \partial_\theta \gamma^{r \varphi}\Big]
      }{
        2\sqrt{1+\zeta_\theta^{n+1}(u_r^{n+1}, u_\theta^{n+1}, u_\varphi^{n+1})} + \sqrt{1+\zeta_\theta^{n+1}(u_r^{n+1}, u_\theta^{n+1}, u_\varphi^{n+1})}
      } 
\Bigg\},
\end{split}
\end{equation}
\end{figure*}

\begin{figure*}[h]
\begin{equation}
\frac{u_\varphi^{n+1}-u_\varphi^n}{\Delta t} = 0\,.
\end{equation}
\end{figure*}
We have defined the following quantities for conciseness:
\begin{align}
\xi_i^{n}
  &:= \gamma^{ii}(u_i^{n})^2
     + 2\gamma^{ij} u_i^{n} u_j^{n}
     + 2\gamma^{ik} u_i^{n} u_k^{n}
     + 2\gamma^{jj} (u_j^{n})^2, 
\\
  &\quad \nonumber
     + 2\gamma^{kk} (u_k^{n})^2
     + 2\gamma^{jk} u_j^{n} u_k^{n},
\\[1ex]
\xi_i^{n+1}
  &:= \gamma^{ii}(u_i^{n+1})^2
     + 2\gamma^{ij} u_i^{n+1} u_j^{n}
     + 2\gamma^{ik} u_i^{n+1} u_k^{n}
     + 2\gamma^{jj} (u_j^{n})^2, 
\\
  &\quad \nonumber
     + 2\gamma^{kk} (u_k^{n})^2
     + 2\gamma^{jk} u_j^{n} u_k^{n}
     \\[1ex]
\zeta_i^{n+1}
  &:= \gamma^{lm}(x^{i, (n+1)}, x^{l, (n+1)}, x^{m, (n+1)}) u_l u_m, 
     \\[1ex]
\zeta_i^{n}
  &:= \gamma^{lm}(x^{i, (n)}, x^{l, (n)}, x^{m, (n)}) u_l u_m.
\end{align}

Note that in Eqs.~\eqref{ur1}--\eqref{ut1} the derivatives of the metric
components are reported analytically, but they are replaced by the
corresponding finite-difference expressions whenever the denominator is
sufficiently large to avoid numerical issues. As an example, for the
derivative $\partial_i \gamma^{lm}$ we use the approximation $\partial_i
\gamma^{lm} \simeq [\gamma^{lm}(x^{i, (n+1)})-\gamma^{lm}(x^{i, (n)})]
/(x^{i, (n+1)}-x^{i, (n)})$. We introduce a threshold $\varepsilon$ on the
coordinate increment $x^{i, (n+1)}-x^{i, (n)}$ and compute the derivative
analytically only when the denominator is smaller than $\varepsilon$, 
which we set to $\varepsilon = 10^{-10}$. The same prescription is
applied to the derivatives $\partial_i \alpha$ and $\partial_i \beta^j$, 
in all the momentum equations.

\end{appendix}
\end{document}